\documentclass[final,12,times,twocolumn]{elsarticle}
  
\journal{New Astronomy}

\begin{document}

\begin{frontmatter}



\title{Massive and refined: a sample of large galaxy clusters simulated at high resolution.
 I:Thermal gas and properties of shock waves.}


%\author{F.Vazza; G.Brunetti; C.Gheller; R.Brunino}
\author{F. Vazza[1,3]; G.Brunetti[1]; C.Gheller[2]; R.Brunino[2]}
\address[1]{INAF/Istituto di Radioastronomia, via Gobetti 101, I-40129 Bologna,
Italy}
\address[2] {CINECA, High Performance System Division, Casalecchio di Reno--Bologna, Italy}
\address[3]{vazza@ira.inaf.it}

\begin{abstract}
We present a sample of 20 massive galaxy clusters with total virial
masses  in the range of $6 \cdot 10^{14}M_{\odot} \leq M_{vir} \leq 2 
\cdot 10^{15}M_{\odot}$, 
re-simulated with a customized
version of the 1.5. ENZO code employing Adaptive Mesh Refinement.
This technique allowed us to obtain
unprecedented high spatial resolution ($\approx 25kpc/h$) up to the distance of 
$\sim 3$ virial radii from the clusters center, and makes it possible to focus
with the same level of detail on the physical properties of the innermost and of the outermost cluster regions, providing new clues on the role of shock waves and turbulent
motions in the ICM, across a wide range of scales. 

In this paper, a
first exploratory study of this data set is presented. We report  
on the thermal properties of galaxy clusters at $z=0$.
Integrated and morphological properties of gas density, gas temperature,
gas entropy and baryon fraction distributions are discussed, and compared with existing outcomes 
both from the observational and from the numerical literature.
 Our cluster sample
shows an overall good consistency with the results obtained adopting other numerical techniques (e.g. Smoothed Particles Hydrodynamics), yet it provides a more 
accurate representation of the accretion patterns far outside the cluster 
cores.
We also reconstruct the properties of shock waves within the sample by means
of a velocity-based approach, and we study
Mach numbers and energy distributions for the various dynamical states in clusters, giving estimates for the injection of Cosmic Rays particles at shocks.
The present sample is rather unique in the panorama of cosmological simulations
of massive galaxy clusters, due to its
dynamical range, statistics of objects and number of time outputs. For this reason, we deploy a public repository of the available
data, accessible via web portal at http://data.cineca.it.

\end{abstract}


\end{frontmatter}


\section{Introduction}
\label{intr}

Simulating the evolution of Cosmological Large Scale Structures of the Universe
is a challenging task. In the last thirty years different
numerical techniques were designed to follow the dynamics of the most important
matter/energy components of the Universe: Dark Matter (DM), baryonic 
matter, and dark energy. 
In order to account for the great complexity and for the number of details provided
by real cluster observations, a number of physical processes in addition
to gravitational collapse and non-radiative hydro-dynamics have been
implemented in many numerical works in the last few years: radiative gas processes, magnetic fields, star formations, AGN feedback, Cosmic Rays, turbulence, etc.(e.g. Dolag et al.2008; Borgani \&
Kravtsov 2009, and references therein, for a recent review). 

At present, two main numerical approaches are massively applied to cosmological
numerical simulations: Lagrangian methods, which sample both the
DM and the gas properties using point-like fluid elements, usually
regarded as particles (e.g.
Smoothed Particles Hydrodynamics codes, SPH) and Eulerian methods, which reconstruct the
gas properties with a discrete space sampling with regular or adaptive meshes
and model the Dark Matter properties with a Particle Mesh approach 
(see Dolag et al.2008 and references therein for a modern review).

\begin{figure*} 
\begin{center}
\includegraphics[width=0.45\textwidth]{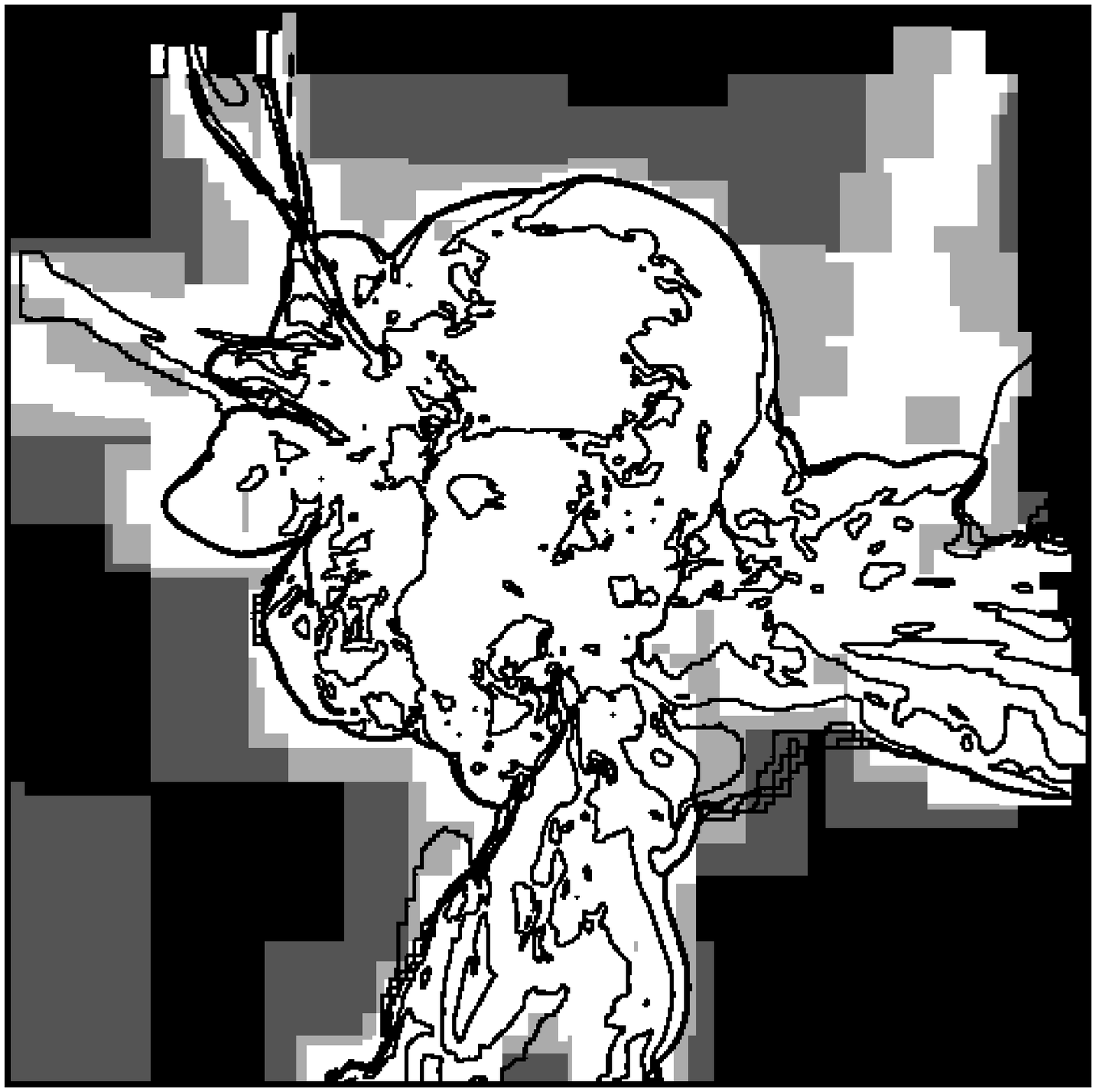}
\includegraphics[width=0.45\textwidth]{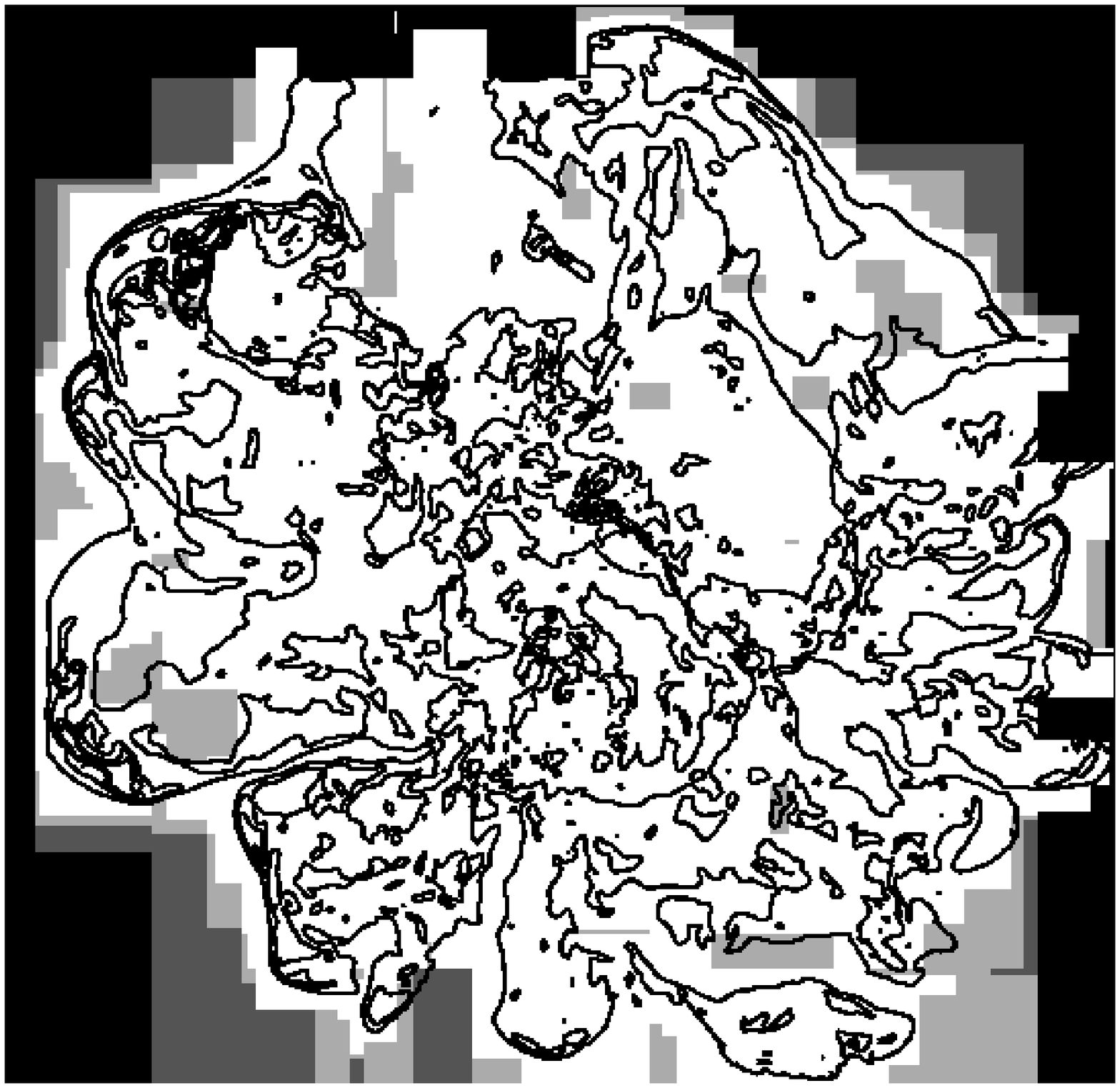}
\caption{The hierarchy of refinement levels in our runs. Color maps: level
of mesh refinement for slices through the center of cluster E1 (left panel)
and E18A (right panel) at $z=0$, from level=0 ($\Delta=200kpc/h$,
in black color) to level=3 ($\Delta=25kpc/h$, in white color); the contour
map shows the gas temperature distribution within the same region (the contours are equally spaced in $\Delta log(T) \approx 0.5$). The side
of both images is $\approx 14Mpc/h$.}
\label{fig:levels}
\end{center}
\end{figure*}

\bigskip

High resolution, AMR simulations (such as the ENZO simulations
presented in this paper) can provide an accurate representation of the
cosmic gas dynamics in galaxy clusters, achieving a very large dynamical
range. Recent works have shown that the adoption of proper mesh refinement
criteria allows to 
study also the details of chaotic motions in the ICM (e.g. Iapichino \& Niemeyer 2008, Vazza et al.2009;
Maier et al.2009; Vazza, Gheller \& Brunetti 2010; Paul et al.2010)

Vazza et al.(2009, hereafter Va09) recently focused on the re-simulation of galaxy
clusters by employing a new mesh refinement criterion, which couples the
``standard'' refinement criteria based on large gas or DM over-densities,
to the mesh refinement criterion based on cell to cell 1--D jumps 
of the velocity field. In Va09 and Vazza, Gheller \& Brunetti (2010, hereafter VGB10) we 
showed that the extra-refinement on 1--D velocity jumps opportunely increases the number of 
resolution elements across the ICM volume, allowing us to achieve a 
better spectral and morphological representation of chaotic motions in the ICM. Furthermore it reduces the artificial dampening of
mixing motions due to the effect of the coarse resolution.

Since the above works were focused on the re-simulation of a few intermediate
mass systems (e.g. $M<3 \cdot 10^{14}M_{\odot}$), it is interesting now
to  extend the same method to a larger sample of higher mass clusters.
Here we present the first results obtained 
analyzing 20 galaxy clusters, with total masses in the range 
$6 \cdot 10^{14}M_{\odot} \leq M_{vir} \leq 2 \cdot 10^{15}M_{\odot}$, 
obtained with the above techniques and designed to
reach very high {\it spatial} resolution around
both DM/gas clumps, shocks and turbulent motions.
Such rich sample accounts for objects of very different dynamical history
and it is characterized by
a large dynamical range ($N_{AMR}\sim 500^{3}$, where $N_{AMR}$ is the number of grid elements 
at the maximum mesh refinement level) within the clusters volume.  
This allows us to study a broad variety of
multi-scale phenomena associated to cluster growth and evolution.

The paper is organized as follows: in Section \ref{sec:methods} we present the
clusters sample, the numerical techniques adopted and the archiving
procedure for the data sample; in Section \ref{subsec:scaling}
we present the integrated (e.g. scaling laws) properties of our clusters,
in Section \ref{subsec:profiles} we present the
and radial properties of gas density, gas temperature and gas entropy
for all clusters in the sample. In Section \ref{subsec:shocks} we characterize
shock waves within the clusters and give estimates on the energy level
of injected Cosmic Rays particles.
The discussion and the conclusions are reported in Section \ref{sec:conclusions}.
In the Appendix, we report consistency tests for the adopted re-ionization
scheme (Sect.\ref{sec:appendix1}), and present a visual inspection of all clusters of the sample (Sect.\ref{sec:appendix2}).

\begin{table}
\label{tab:clusters}
\caption{Main characteristics of the simulated clusters at $z=0$. 
Column 1: identification number; 2:
total virial mass ($M_{vir}=M_{\rm DM}+M_{gas}$); 3: virial radius ($R_{v}$); 4:$X=E_{k}/E_{tot}$ ratio inside $R_{v}$; 5:dynamical classification: RE=relaxing, ME=merging or  MM=major merger (with approximate redshift of  the last merger event).}
%\begin{center}
\begin{tabular}{c|c|c|c|c|c|c|c}
ID & $M_{vir}$ & $R_{v}$ & $X$  & note  \\
   & [$10^{15}M_{\odot}$] & [$Mpc$] & [$E_{kin}/E_{tot}$] & \\
\hline

E1 & 1.12  & 2.67 & 0.43  & MM(0.1)\\
E2 & 1.12 & 2.73 & 0.47  & ME  \\
E3A & 1.38 & 2.82 & 0.43 & MM(0.2) \\
E3B & 0.76 & 2.31 & 0.55 & ME \\
E4 & 1.36 & 2.80 & 0.44 & MM(0.5)\\   
E5A & 0.86 & 2.39 & 0.47 & ME \\  
E5B & 0.66 & 2.18  & 0.75  & ME \\
E7 & 0.65 & 2.19  &  0.45  & ME \\ 
E11 & 1.25 & 2.72  & 0.40  & MM(0.6)\\ 
E14 & 1.00 & 2.60 & 0.23 & RE\\
E15A & 1.01 & 2.63 & 0.85   & ME\\
E15B & 0.80 & 2.36 & 0.33 & RE\\
E16A & 1.92 & 3.14  & 0.36  & RE \\ 
E16B & 1.90 & 3.14  & 0.67  & MM(0.2) \\ 
E18A & 1.91 & 3.14  & 0.37 & MM(0.8) \\  
E18B & 1.37 & 2.80  & 0.34 & MM(0.5)\\   
E18C & 0.60  & 2.08  & 0.55 & MM(0.3) \\  
E21 & 0.68 & 2.18 & 0.40 & RE\\ 
E26 & 0.74 & 2.27 & 0.29 & MM(0.1)\\  
E62 & 1.00 & 2.50  & 0.63 & MM(0.9) \\
\end{tabular}
\label{tab:char}
%\end{center}
\end{table}

\section{Numerical Methods}
\label{sec:methods}

\subsection{The ENZO code}

Computations presented in this work were performed using the 
ENZO code developed by the Laboratory for Computational
 Astrophysics at the University of California in San Diego 
(http://lca.ucsd.edu).

ENZO is an adaptive mesh refinement (AMR) cosmological hybrid 
code highly optimized for high performance computing
(Bryan \& Norman 1997; Norman et al.2007). 

It uses a particle-mesh N-body method (PM) to follow
the dynamics of the collision-less Dark Matter (DM) component 
(Hockney \& Eastwood 1981), and an adaptive mesh method for ideal 
fluid-dynamics (Berger \& Colella, 1989).

The DM component is coupled to
the baryonic matter (gas), via gravitational forces, calculated from the 
total mass distribution (DM+gas) solving the Poisson equation with 
a FFT based approach.
The gas component is described as a perfect fluid and its dynamics
is calculated solving conservation equations of mass, energy and momentum over a computational mesh,
using an Eulerian solver based on the 
Piecewise Parabolic 
Method (PPM, Woodward \& Colella, 1984). This scheme is 
a higher order extension of Godunov's shock capturing
method (Godunov 1959), and it is at least second--order accurate in space outside
of shocks (up
to the fourth--order in 1--D, in the case of smooth flows and small time-steps) and
second--order accurate in time.

\begin{figure*} 
\begin{center}
\includegraphics[height=0.98\textwidth,angle=90]{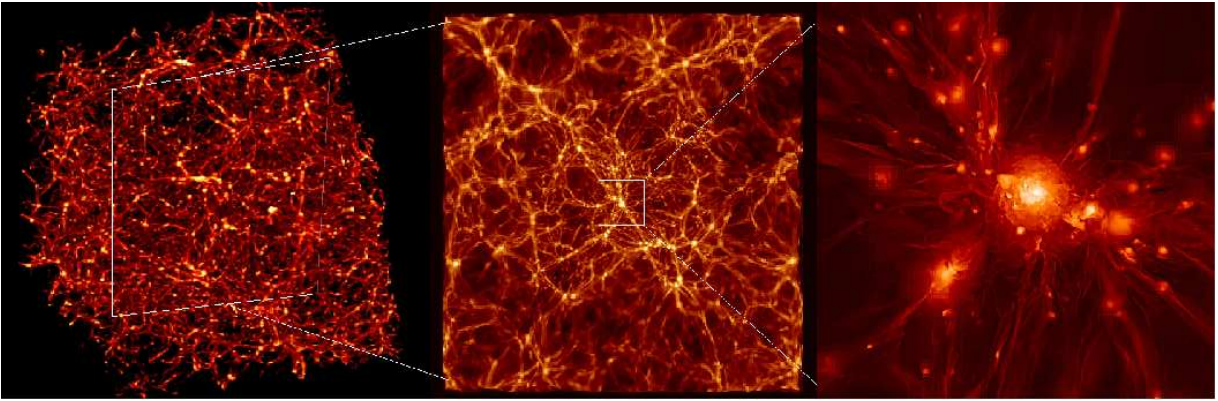}
\caption{{\it Left:} 3-D rendering of gas matter within a $187Mpc/h$ computational box at $z=0$. {\it Center:} slice of thickness $\approx 20Mpc/h$ and with the size of $187Mpc/h$ centered in the AMR region.  The renderings 
are done with Visivo (Comparato et al.2007, http://visivo.cineca.it).{\it Right:} 3-D distribution
of Dark Matter density inside the AMR region (the side of the image is $\approx 8Mpc/h$) at $z=0$.}
\label{fig:LLS}
\end{center}
\end{figure*}

\subsection{Clusters simulations}
\label{subsec:amr}

For the simulations presented here, we assumed a ``concordance'' $\Lambda$CDM cosmology with
$\Omega_0 = 1.0$, $\Omega_{BM} = 0.0441$, $\Omega_{DM} =
0.2139$, $\Omega_{\Lambda} = 0.742$, Hubble parameter $h = 0.72$ and
a 
normalization for the primordial density power
spectrum $\sigma_{8} = 0.8$.

The clusters considered in this paper were extracted
from a few simulations of cosmological volumes with linear size of 
$\approx 190Mpc/h$.
For each simulation we adopted a root grid of $220^{3}$ cells and the same number of DM particles. This leads to a
DM mass resolution of 
$m_{dm} \approx 4.3 \cdot 10^{10} M_{\odot}/h$. The overall simulated cosmic volume 
is $\sim 480^{3} (Mpc/h)^{3}$ for the whole cluster sample.
This initial set of simulations was used to find the most massive clusters
in the volume, targets of the re-simulations at higher spatial
and mass resolution.

The most massive objects of all boxes where identified with an halo finder
algorithm working on gas/DM spherical over-density (e.g. Gheller, Pantano \& Moscardini
1998). 
Nested initial
conditions were then applied to the volumes comprising the formation regions 
of all clusters, to achieve a higher DM mass resolution for the cosmic 
volume involving the formation of the clusters. 
In details, two levels of nested initial conditions were placed in cubic
regions centered on the cluster centers. After a few tests, we adopted
the combination of grid sizes which resulted to be the best compromise between
the computational cost and the need for the best possible
resolution in the cluster volumes.

The first level box had the size 
of $\approx 95Mpc/h$ (with $m_{dm} \approx 5.4 \cdot 10^{9} M_{\odot}/h$ 
and constant spatial resolution of $\Delta_{1} \approx 425kpc/h$). The second had a size
of $\approx 47.5Mpc/h$ (with $m_{dm} \sim 6.7 \cdot 10^{8}M_{\odot}/h$ and 
constant spatial resolution of $\Delta_{1} \approx 212kpc/h$).
For every cluster run, we identified cubic regions
with the size of $\sim 6 R_{v}$ (where
$R_{v}$ is the virial
radius of clusters at $z=0$, calculated on lower resolution fast runs), 
and allowed the code to apply 3 additional levels of mesh refinement, 
achieving a peak spatial resolution of $\Delta \approx 25kpc/h$; 
in the following, we will refer to this sub-volume as to the ``AMR region''.

From $z=30$ (initial redshift of the simulation) to $z=2$, mesh refinement is 
triggered by gas or DM over-density criteria. From $z=2$ an additional refinement 
criterion based on 1--D velocity
jumps (Va09) is switched on. This
second AMR criterion is designed to  
capture shocks and intense turbulent motions in the 
ICM out to the clusters outskirts. The reader can refer to
Va09 and VGB10, where we presented a detailed comparison of the differences
in the properties of thermal gas, shocks and turbulence distributions found
when comparing the standard and the our extended mesh refinement method. 

Compared to the standard mesh refinement strategy, 
we showed that
the use of the additional refinement on velocity jumps leads to: 

\begin{itemize}
\item a sharper 
reconstruction of accretion and merger shocks in the clusters
volume; 
\item  a $\sim 10$ percent 
lower gas density and a $\sim 10$ per cent larger gas entropy profile inside
the clusters core; 
\item a substantially enhanced presence of turbulent motions at all radii (up to a factor $\sim 2$ in energy); 
\item a more efficient mixing of gas matter during the whole cluster evolution.
\end{itemize}

All these results are well converged for re-simulations adopting a threshold value of $\delta v/v \leq 3$ to trigger the mesh refinement locally, where $\delta v$ is the
1--D velocity difference and $v$ is the minimum velocity for the cell in the patch of cells to be refined (converge tests can be found in Va09).

We also note that recent works (e.g. Agertz et al.2007; Wadsley et al.2008; Springel 2010; Robertson et al.2010) 
suggested that Eulerian codes can be subject to considerable 
un-physical numerical diffusion, that may lead to a suppression of 
fluid instabilities, in the regions where the mesh resolution is too low
and large bulk motions are present.
However, the opportune triggering of
mesh resolution in these regions, such that implemented in our cluster runs, can overcame this problem (Robertson et al.2010).

\bigskip

In the cluster runs presented here, the number of cells refined up to the 
peak resolution ($25kpc/h$) at $z=0$
varies from  $\sim 20$ to $\sim 40$ per cent of the total volume within 
the AMR region ($N \sim 10^{7}-10^{8}$  cells). 
In Fig.\ref{fig:levels} we show the map of the spatial distribution for the refinement levels, with overlaid contours of gas temperature, for two representative clusters of the sample.

 Approximately, the number of high resolution DM particles 
contained within the AMR region at $z=0$ is of the order of $N_{dm} \sim 2-3 \cdot 10^{7}$, and only a few ($<100$) DM particles coming from the
lower resolution regions are found (but never within the virial volume
of clusters). However, the gravitational
potential in the PM approach is computed after interpolating the DM mass
distribution onto a grid, and no problems of contamination (leading,
for instance, to a spurious transfer of kinetic energy) are present.

Figure \ref{fig:LLS} shows a rendering of the 3--D distribution of gas matter within
the whole computational region of side $187Mpc/h$, and a zoom into the sub-volume
of $\approx 13.6Mpc/h$ of the AMR region for one of the cluster run.

Our runs neglect radiative cooling, star formation processes and AGN-feedback. 
Re-heating due to stars and AGN activity is treated at run-time with 
a simplified approach reproducing an Haardt \& Madau (1996) re-ionization model. A detailed description is reported in the Appendix \ref{sec:appendix1}.

\bigskip

Figure \ref{fig:amr_e18} shows a slice in gas temperature for the biggest
cluster in our sample (E18A), giving the visual
impression of the extraordinary amount of details characterizing
each simulated cluster at $z=0$: gas substructures, sharp shock discontinuities and various kinds of fluid instabilities (e.g. Kelvin-Helmoltz and
Rayleigh-Taylor) can be easily found at all distances from the
cluster core to the most peripheral regions, with similar
resolution.

\bigskip

Approximately, every cluster run took $\sim 30 000$ cpu hours on a linux SP6
cluster at CINECA (Casalecchio di Reno, Bologna), for a total amount
of $\sim 8 \cdot 10^{5}$ hours of CPU time.
One of the future goal of this project is to apply
tracer particles in the study in the ICM. To make this possible in a post-processing
phase (as in VGB10), for every cluster we saved a large number
of time outputs (between $60$ and $90$),
with an approximate time sampling of $0.1Gyr$ for $z<1.0$.   
This huge amount of data will allow the useres for a number of iterative
studies (e.g. focusing on Cosmic Rays injection and advection
in the ICM),  without having to run 
the simulations again.

\begin{figure*} 
\begin{center}
\includegraphics[width=0.99\textwidth]{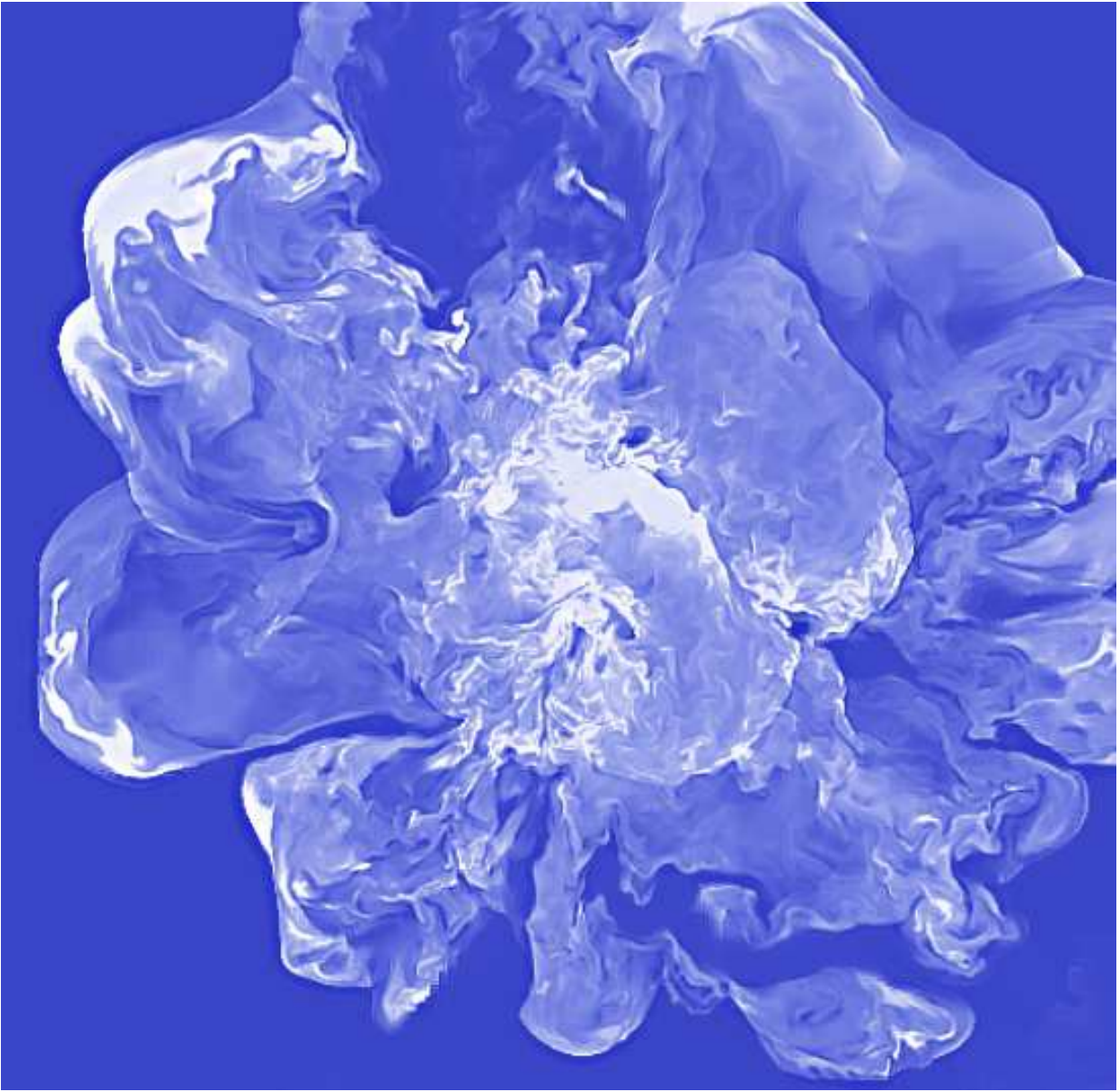}
\caption{A slice in gas temperature for the most massive cluster
in our sample, E18A, at $z=0$. The resolution of the image is $25kpc/h$ per pixel, the size of the image is $13.6Mpc/h$. The underlying structure of mesh
refinements levels is sketched in the right panel of Fig.\ref{fig:levels}.}
\label{fig:amr_e18}
\end{center}
\end{figure*}

\begin{figure*} 
\begin{center}
\includegraphics[width=0.48\textwidth,height=0.42\textwidth]{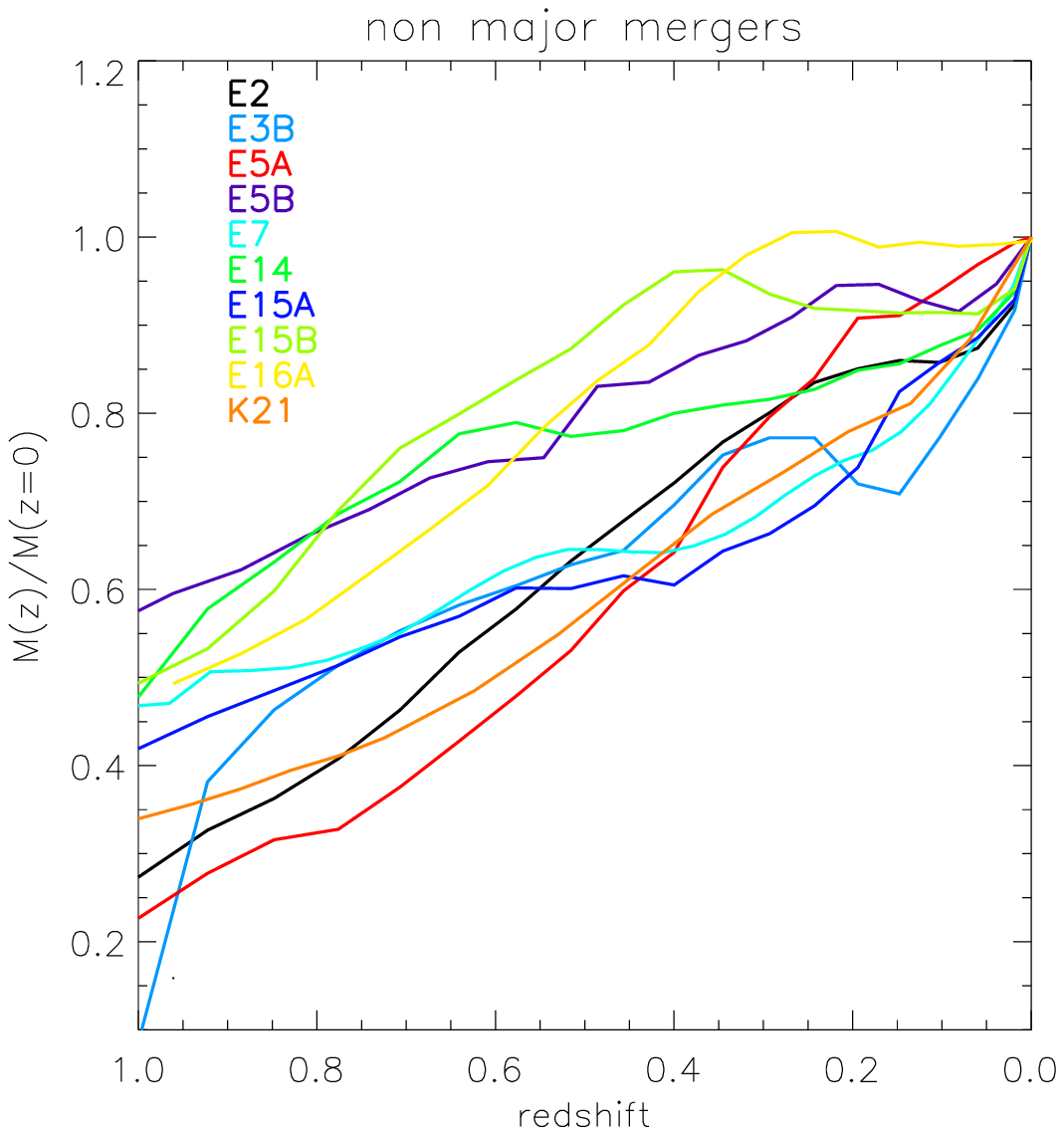}
\includegraphics[width=0.48\textwidth,height=0.42\textwidth]{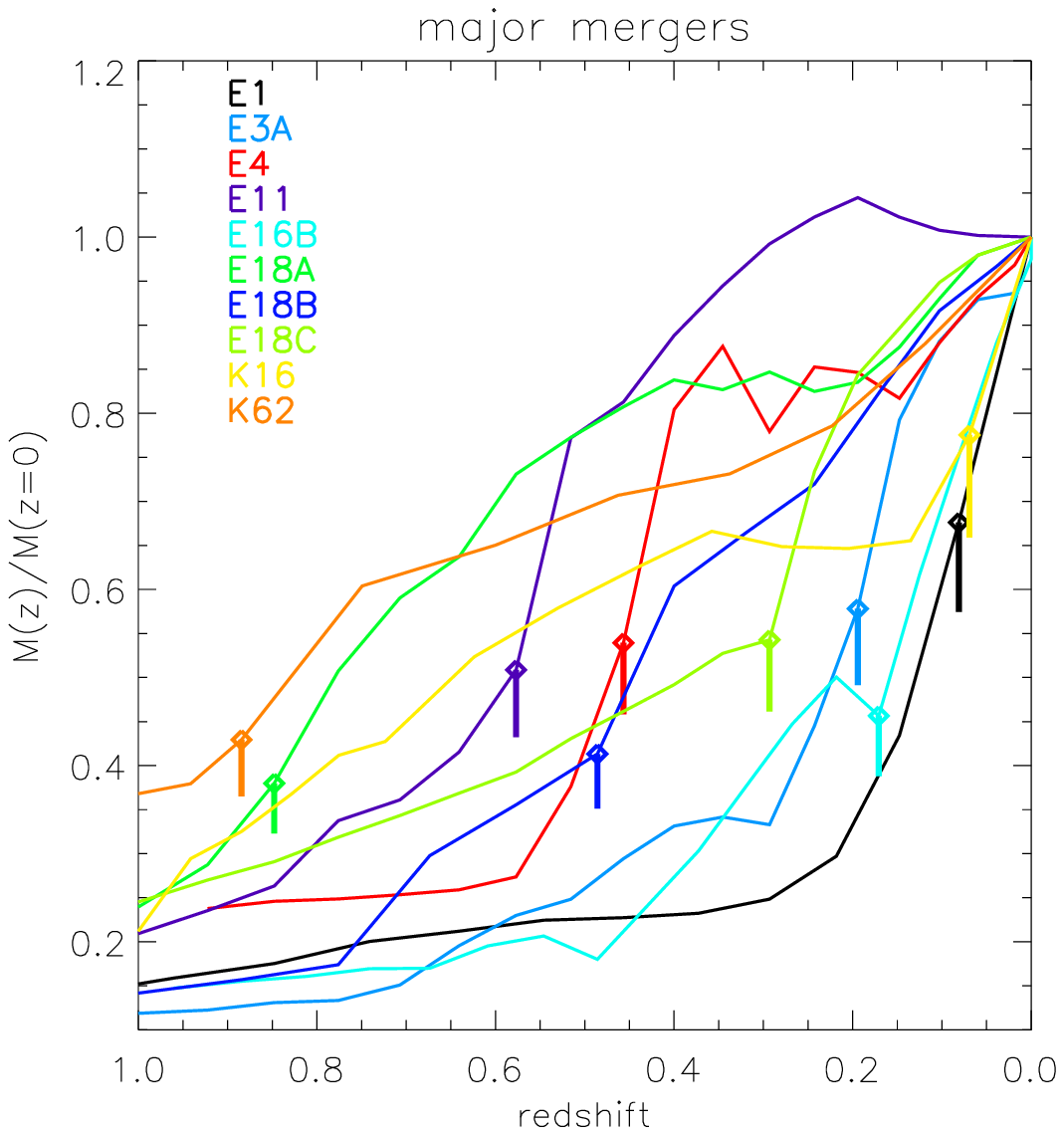}
\caption{Evolution of the total virial mass for all clusters in the sample, normalized for the total mass at $z=0$. The {\it left} panel shows the evolution for {\it non-major merger} clusters, while the {\it right} panel shows the evolution for {\it major merger} clusters; additional arrows show the approximate epoch of the last major merger event for every object.}
\label{fig:matter_evol}
%\end{center}
\end{center}
\end{figure*}

\begin{figure} 
\begin{center}
\includegraphics[width=0.48\textwidth]{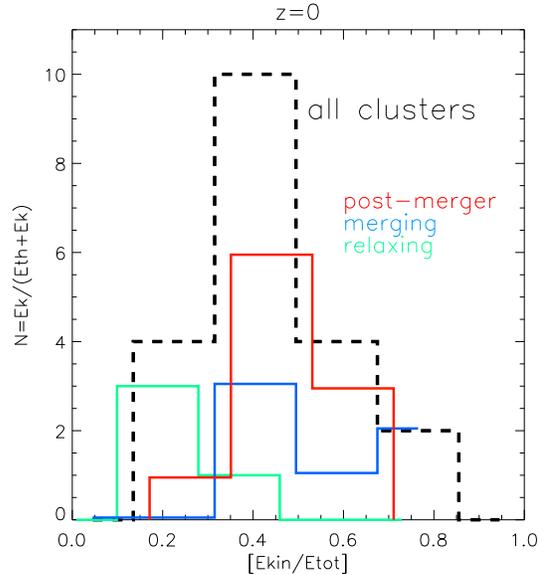}
\caption{Distribution for the ratio between the total kinetic energy, $E_{kin}$
and the total energy (thermal+kinetic), $E_{tot}$, within the virial
volume each cluster in the sample (dashed line). In colors, we additionally
show the distribution of the energy ratio for the three classes of 
clusters discussed in the paper.}
\label{fig:kin}
%\end{center}
\end{center}
\end{figure}

\begin{figure*} 
\begin{center}
\includegraphics[width=0.33\textwidth]{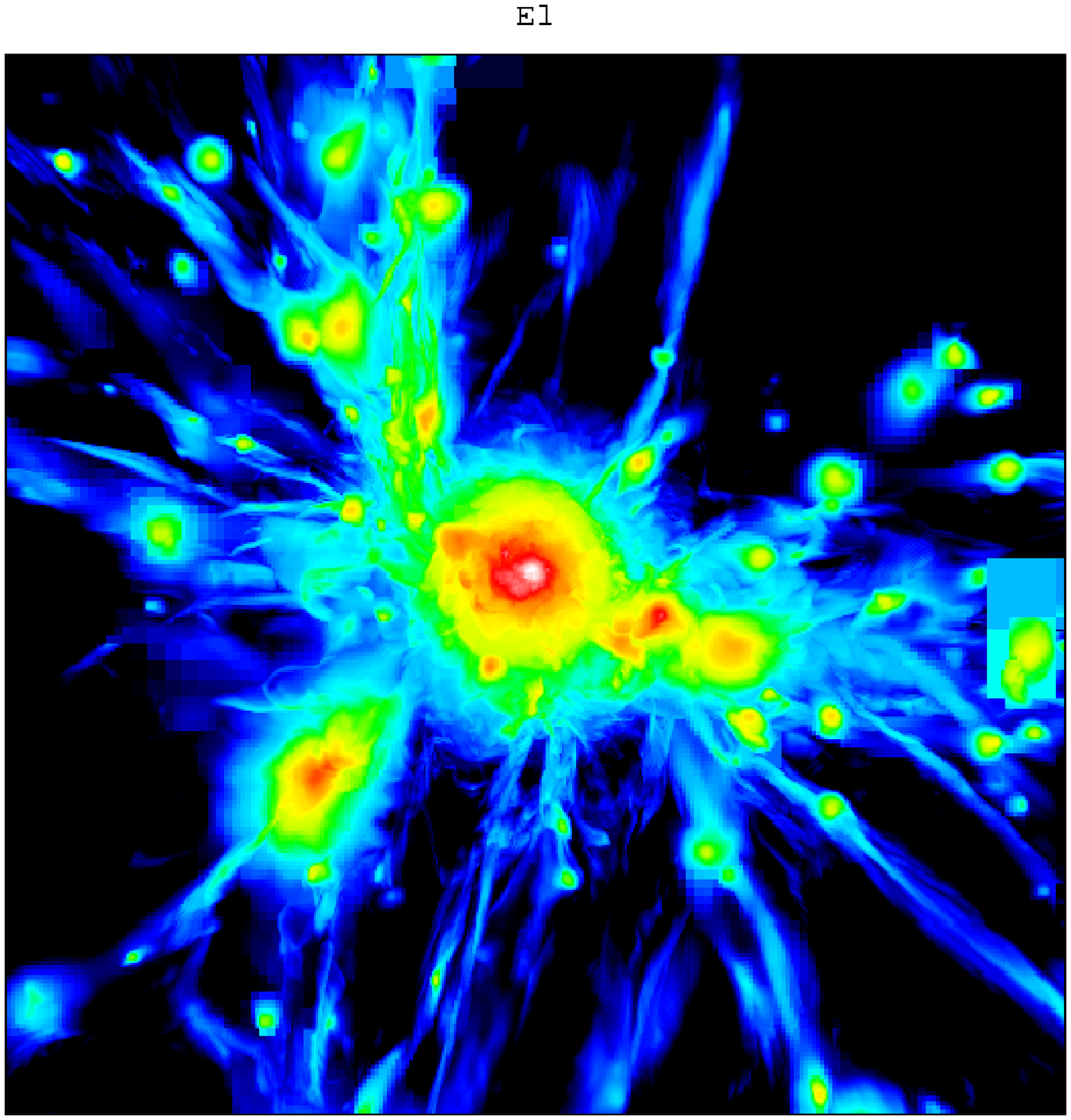}
\includegraphics[width=0.33\textwidth]{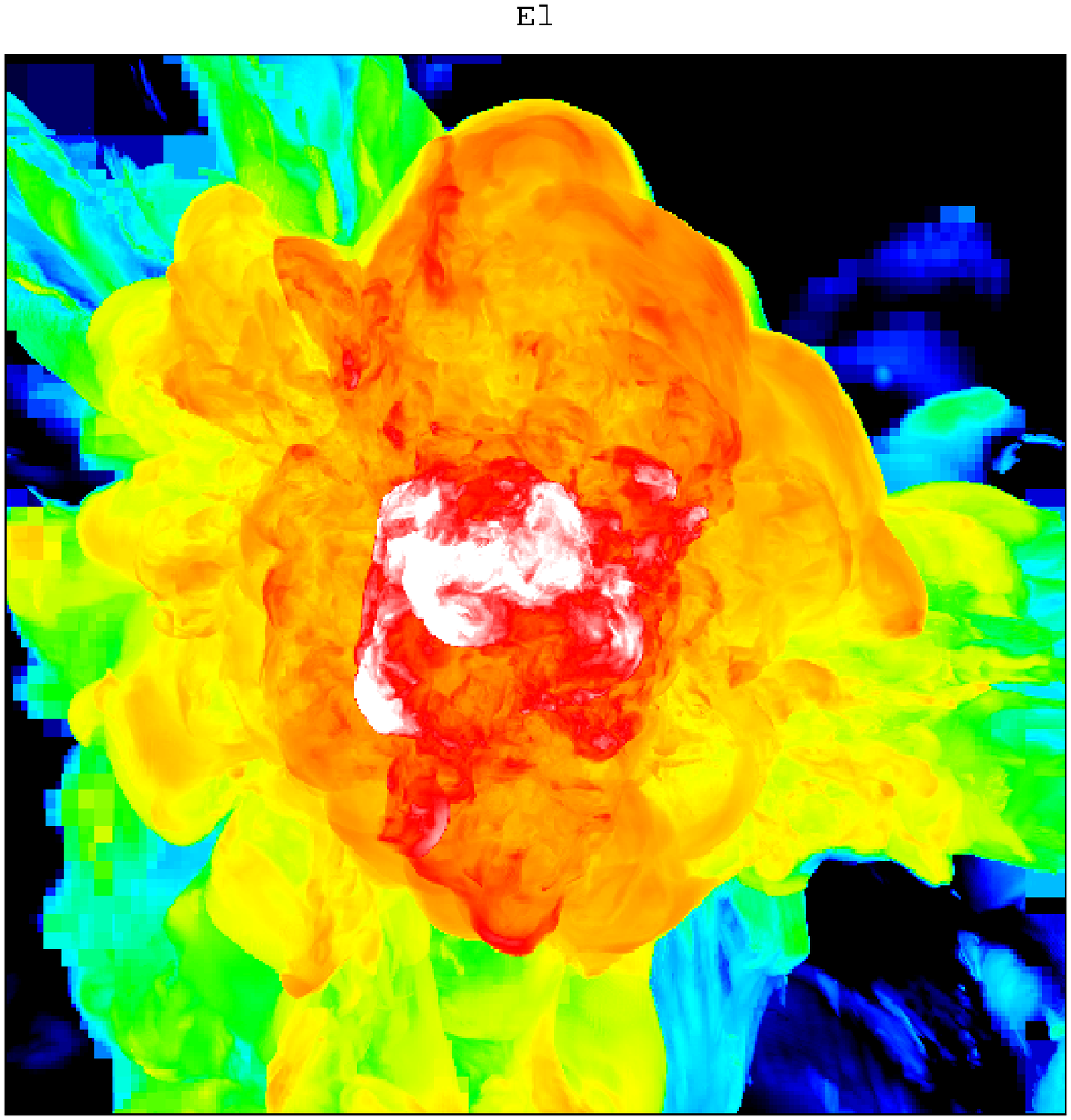}
\includegraphics[width=0.33\textwidth]{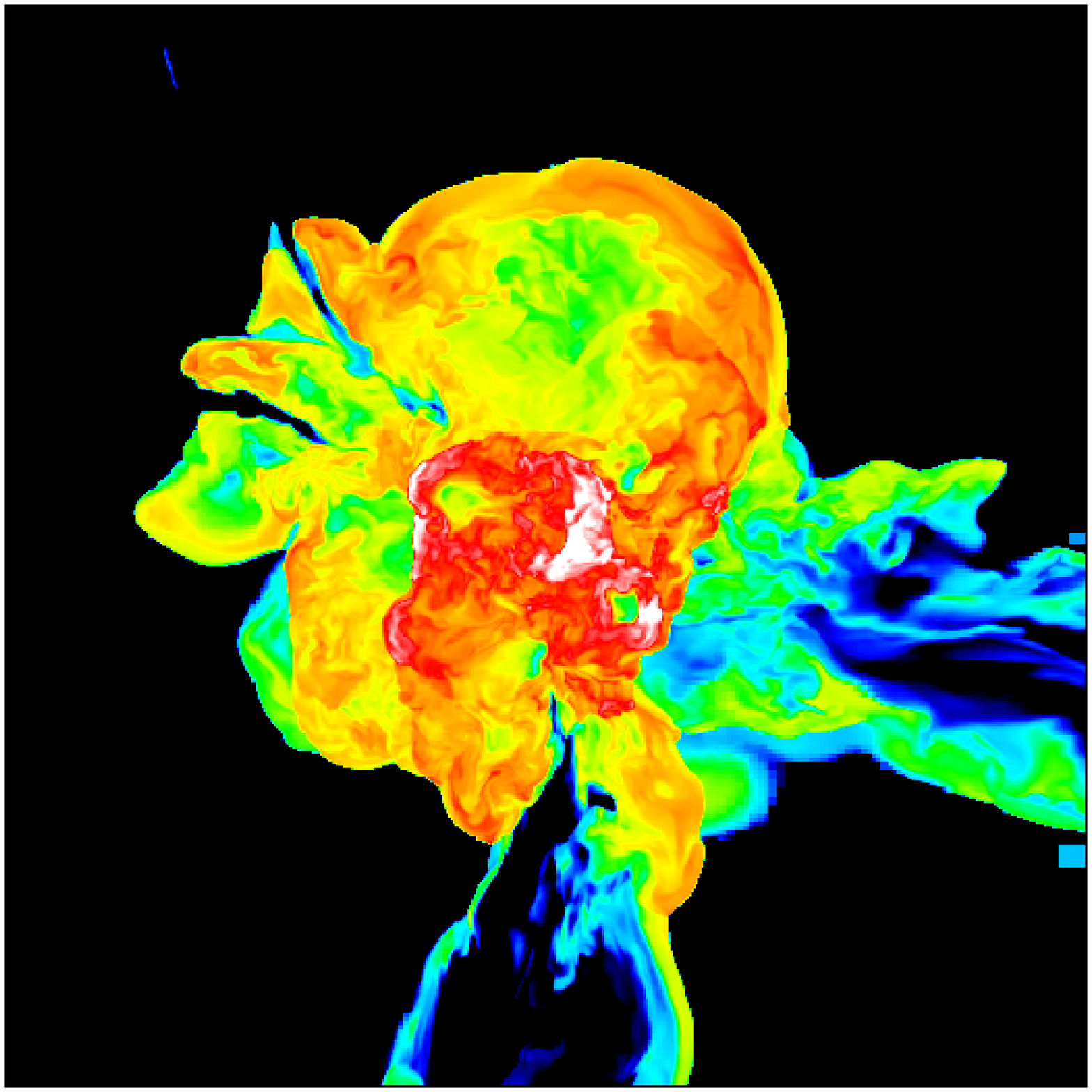}
\includegraphics[width=0.33\textwidth]{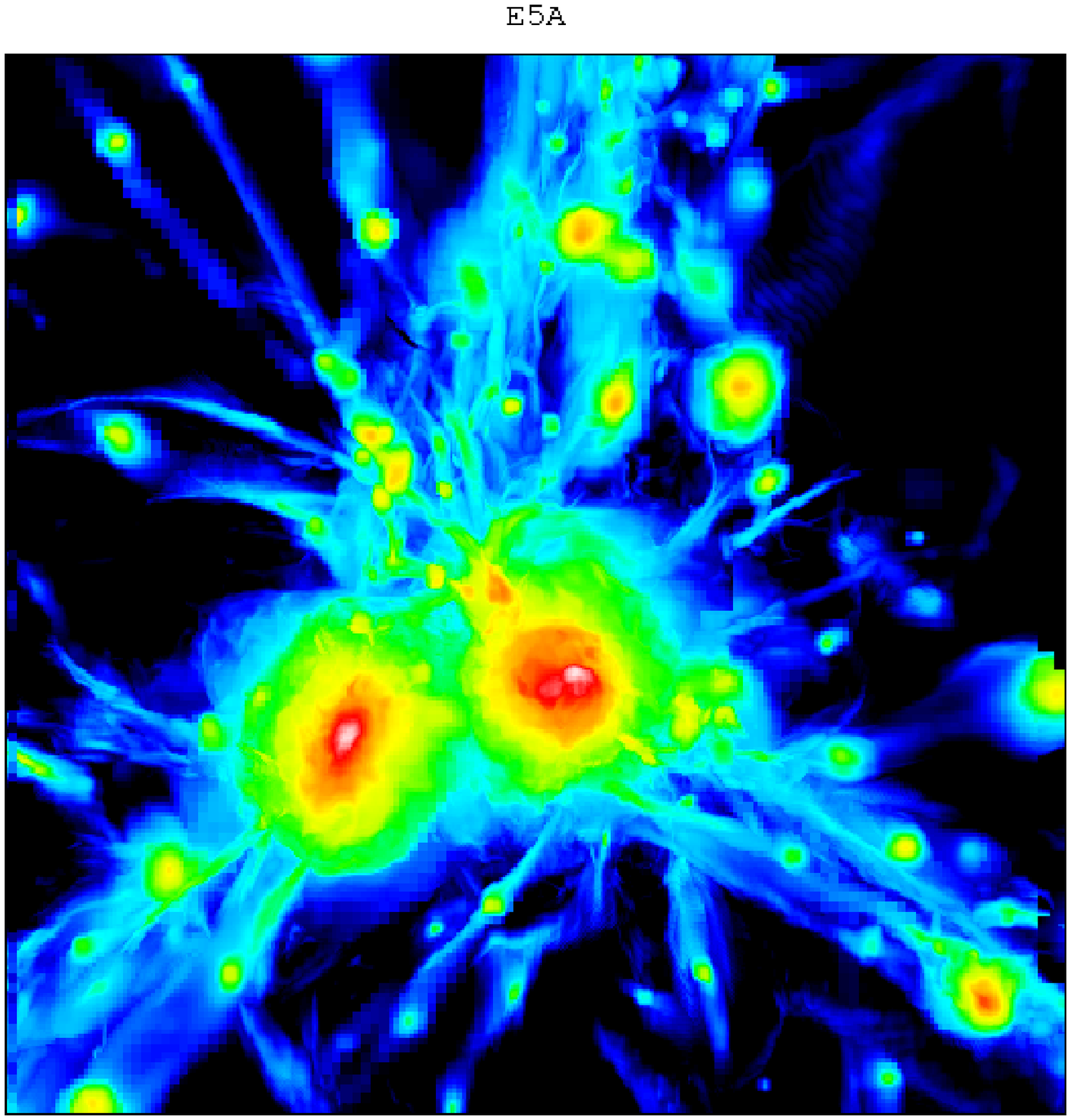}
\includegraphics[width=0.33\textwidth]{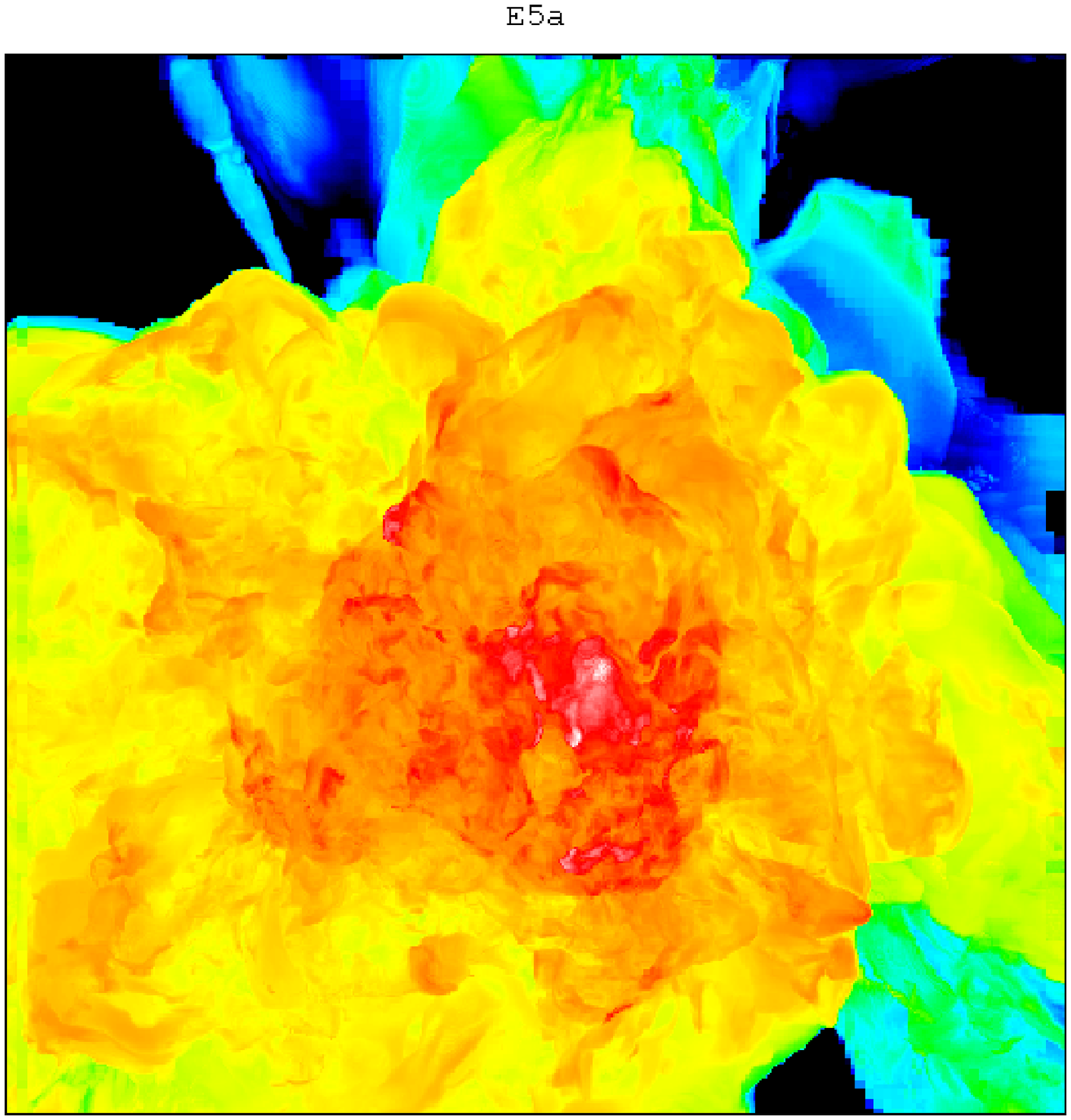}
\includegraphics[width=0.33\textwidth]{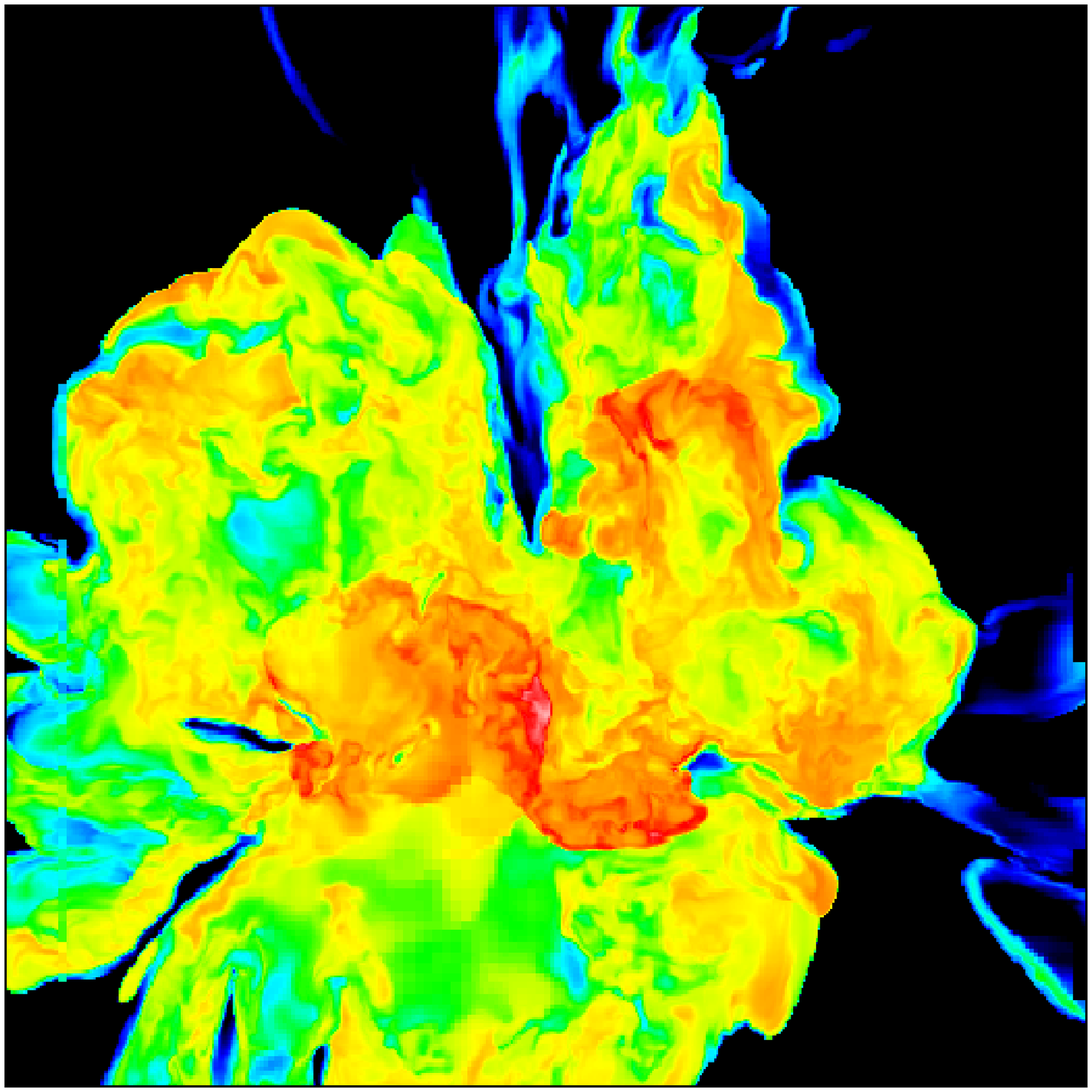}
\includegraphics[width=0.33\textwidth]{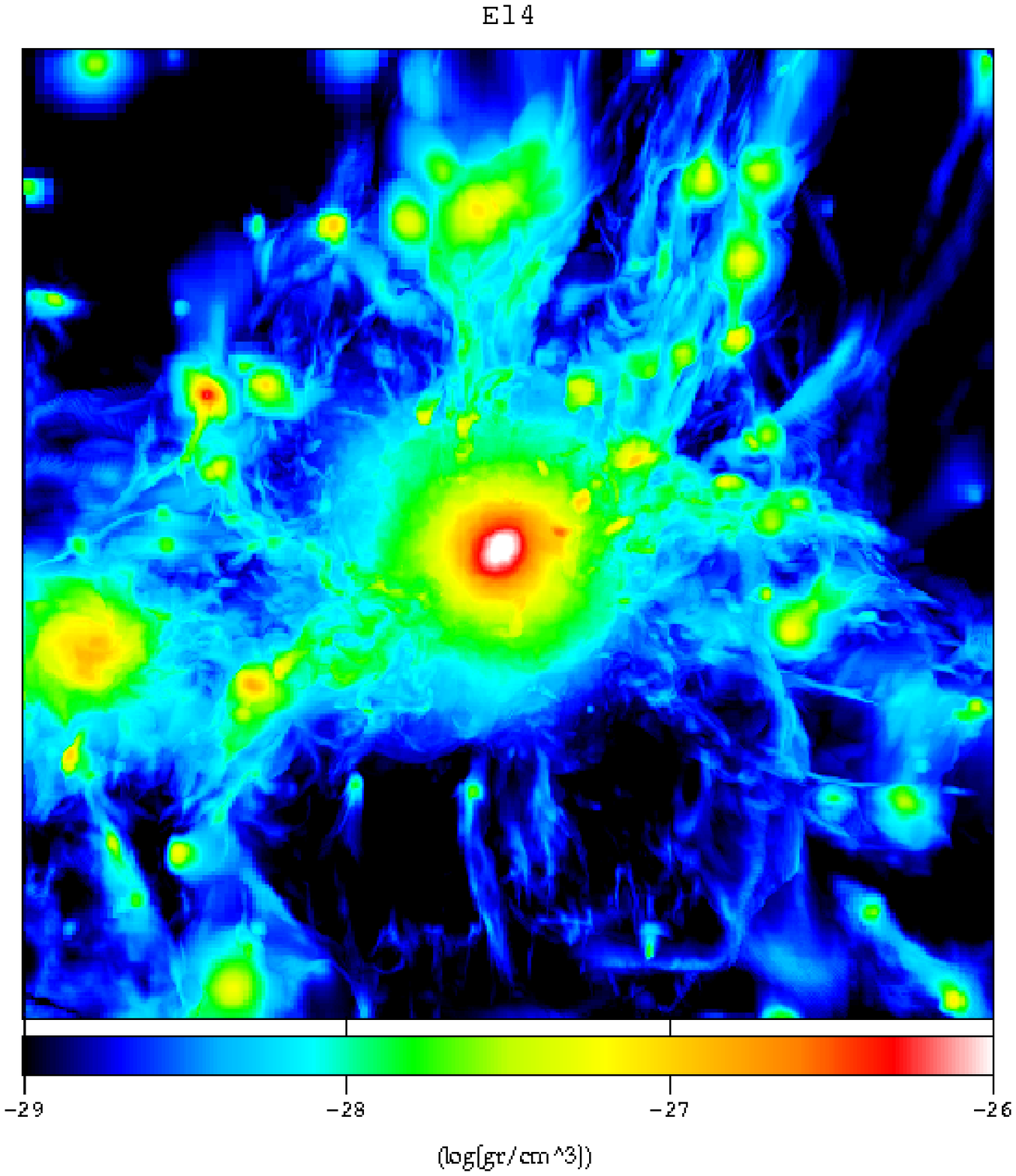}
\includegraphics[width=0.33\textwidth]{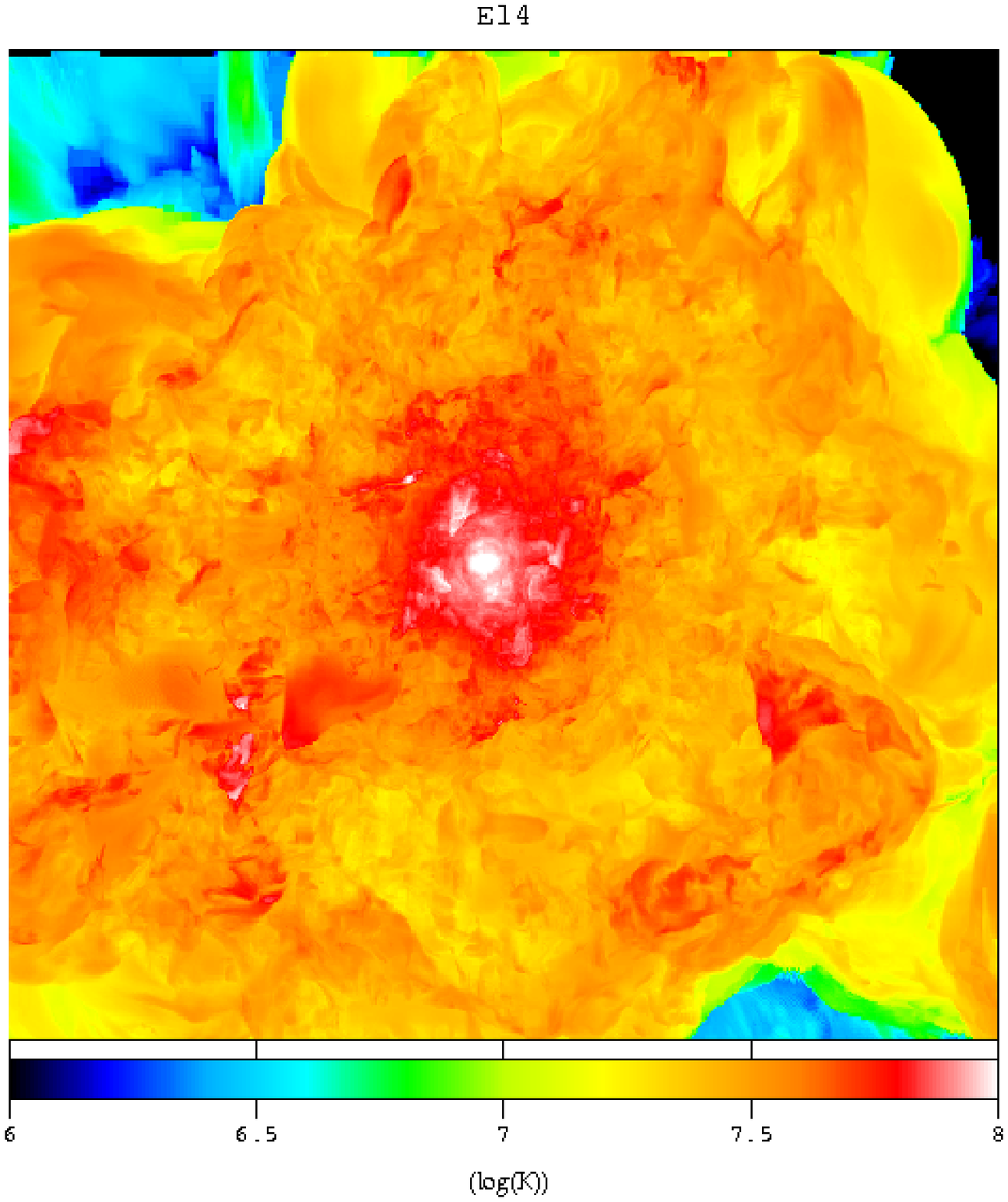}
\includegraphics[width=0.33\textwidth]{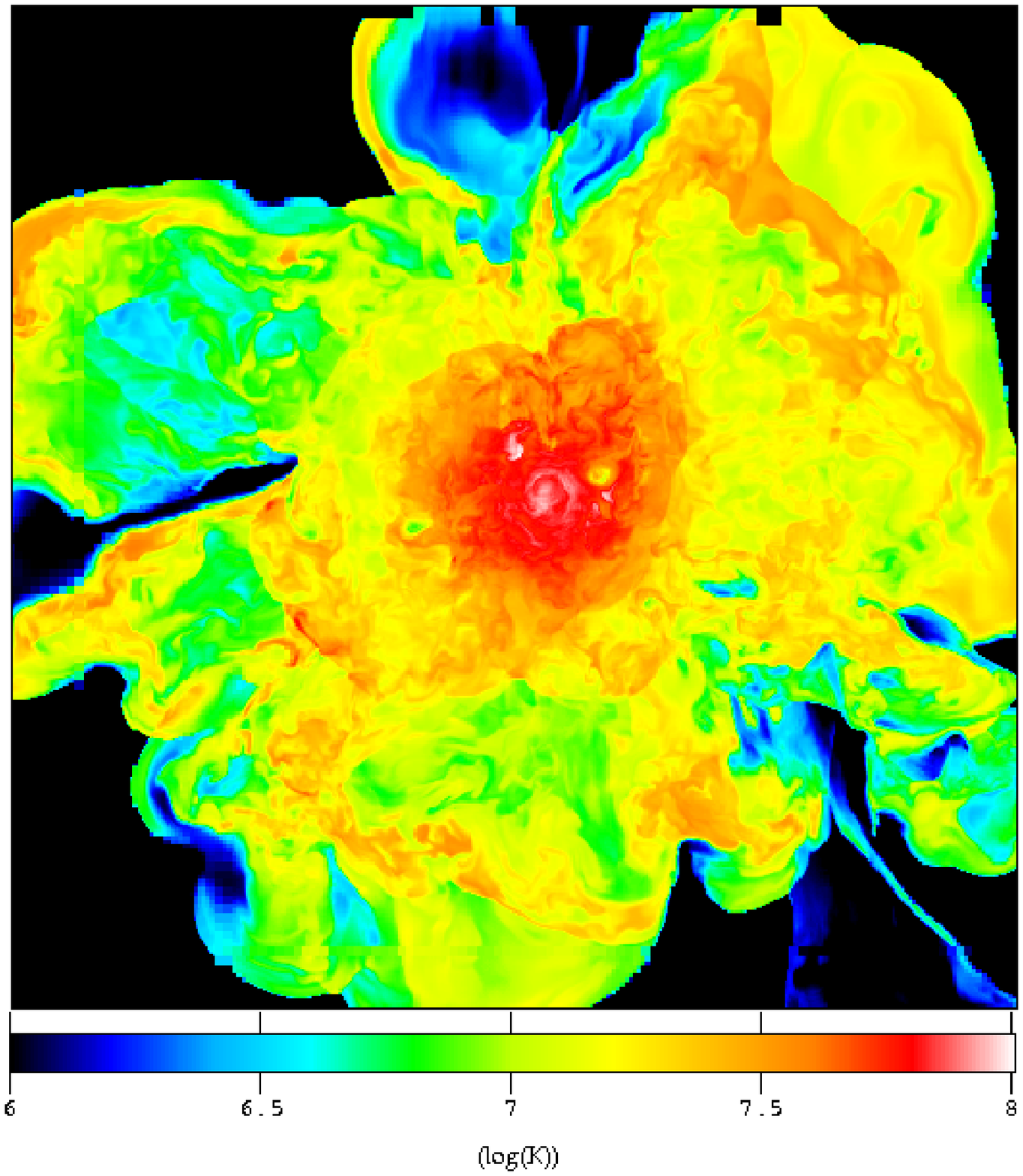} 
\caption{The visual appearance of the three categories of clusters considered in the paper. {\it Top} panels: the major merger cluster E1; {\it center} panels: the merging cluster E5A; {\it bottom} panels: the relaxing cluster E14. From right to left, shown are: maximum gas density along the line of sight ({\it left column}); maximum gas temperature along the line of sight ({\it center column}) and gas temperature in a slice of depth $25kpc/h$ ({\it right column}). The side of each image is $\approx 13.6Mpc/h$.}
\label{fig:categories}
\end{center}
\end{figure*}

\begin{figure*} 
\begin{center}
\includegraphics[width=0.33\textwidth]{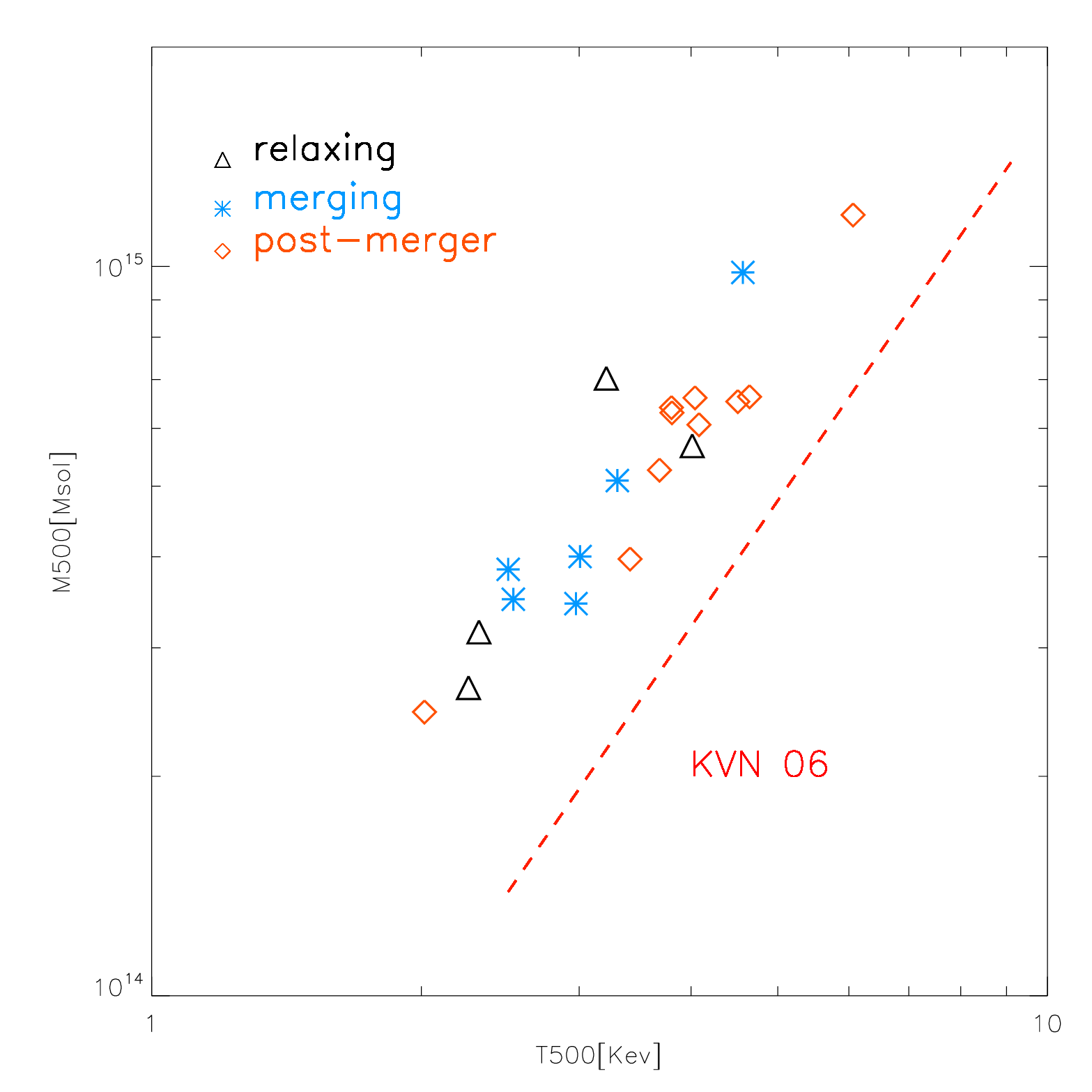}
\includegraphics[width=0.33\textwidth]{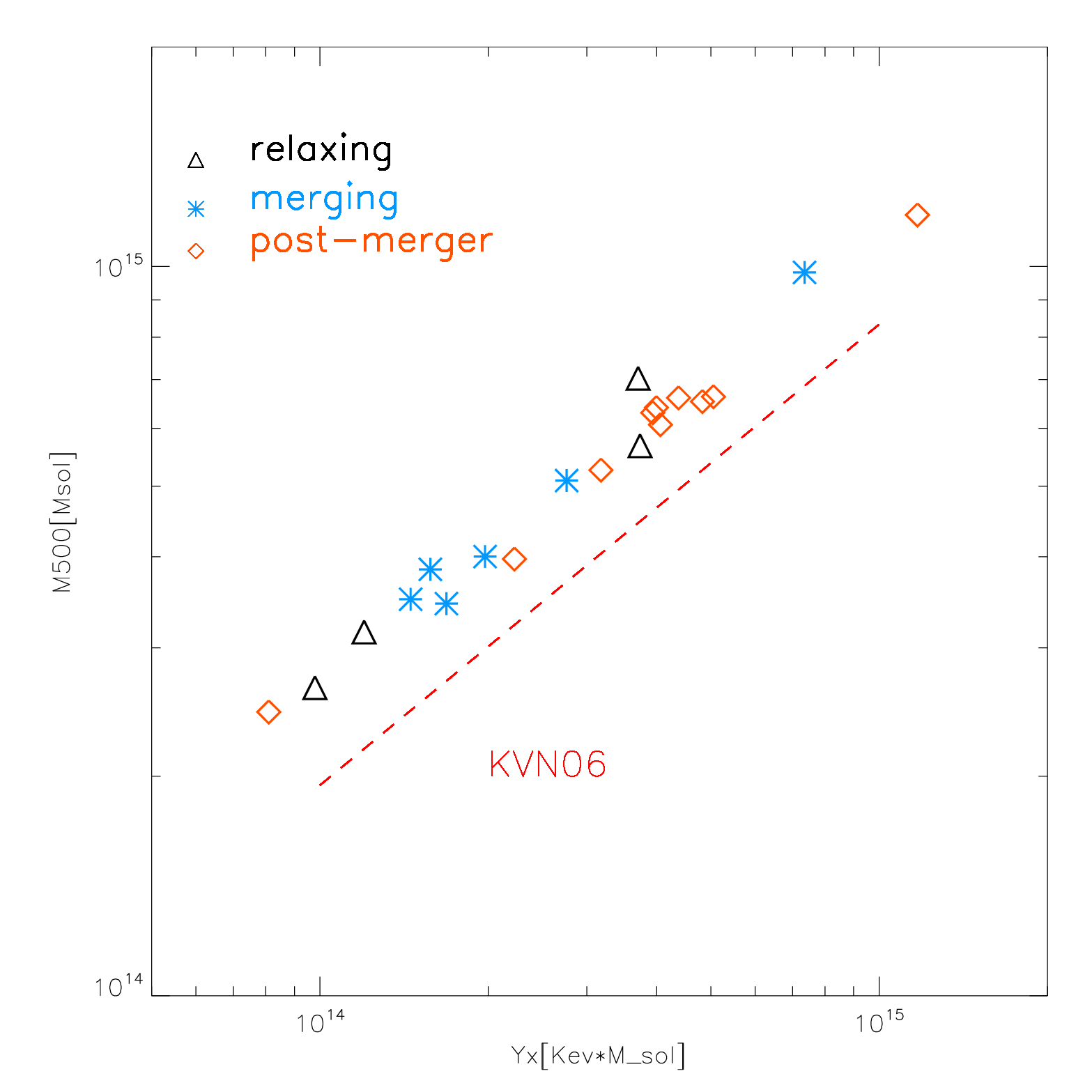}
\includegraphics[width=0.33\textwidth]{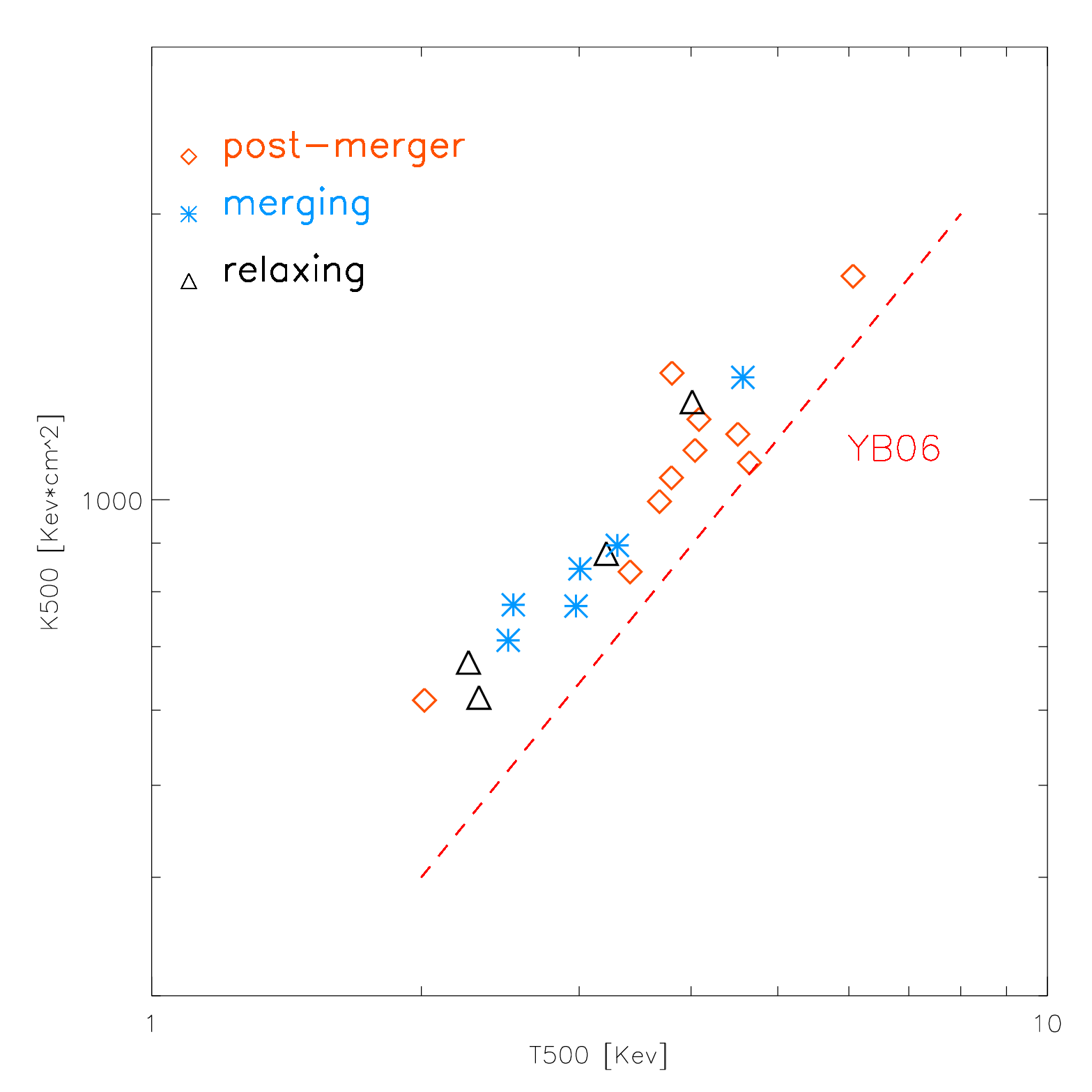}
\caption{Scaling relations for all clusters in the sample at $z=0$. {\it Left
panel}: $M_{500}$ versus $T_{500}$ relation, the additional dashed line
is for the fit relation reported in Kravtsov, Vikhlinin \& Nagai (2006).
{\it Central panel}: $M_{500}$ versus $Y_{500}$ (gas presudo-pressure) relation,
the additional red line shows the ``perfect-slope'' scaling (Kravtsov, Vikhlinin \& Nagai 2006). {\it Right panel}: $S_{500}$ (gas entropy) versus $T_{500}$
relation, the additional red line shows the scaling obtained by Younger \& Bryan (2006).} 
\label{fig:scalings}
\end{center}
\end{figure*}

\subsection{The clusters sample}
\label{subsec:clusters}

Table \ref{tab:clusters} lists all the simulated clusters, along with their main parameters described here below.
All the objects have a total mass $M> 6 \cdot 10^{14}M_{\odot}$,
12 of them having a total mass $M>10^{15}M_{\odot}$. 

This makes our simulated sample a unique tool to study the evolution of the richest cluster of
galaxies in the Universe, in an evolving cosmological framework.

In order to classify the clusters according to their dynamical
state, we adopted two independent proxies
computed for each cluster, in order to account for their main dynamical
differences at $z=0$.

\bigskip 

First, we followed in detail the matter accretion history of all clusters,
in the range $0 \leq z \leq 1$. In Fig.\ref{fig:matter_evol} we show the total (DM+gas) mass evolution  
for every objects, as reconstructed using
the lower mesh refinement level in the AMR region ($\Delta \approx 200kpc/h$). 
The clusters can be broadly grouped into two categories depending on the presence
of a major merger event for $z<1$ epochs (in the 
following we will conventionally use the terms of ``major merger'' or 
``relaxing'' cluster for the two categories). 

A major merger is defined as a
total matter accretion episode where $M(t_{2})/M(t_{1})-1>1/3$ (e.g. Fakhouri, Ma \&
Boylan-Kolchin 2010); 
in our case, we fixed $t_{2}=t_{1}+1Gyr$. The time resolution of $1Gyr$ is of 
the order of the crossing time within the virial volume of clusters, and 
our tests showed that it is small enough to capture the sharp increase of 
cluster mass during strong mergers. 
 
In the last column of Tab.\ref{tab:clusters} we report the approximate epoch
of the last major merger event for major merger objects of the sample; this
procedure divides our total sample in two equal classes of 10 objects each.

We noted, however, that the analysis of the matter accretion history 
for the virial
volume is not sensitive enough to account for the possible variety of
cluster morphologies. This is the case of clusters
experiencing the early stage of a strong merger event (i.e. ``merging'' 
systems), that can
make their overall morphology significantly asymmetric and perturbed
even if the total mass has not yet increased in a significant way. 

For this reason, we computed a second, independent proxy of the dynamical 
state of each cluster, measuring the ratio between the total kinetic energy of 
gas motions inside the virial region, $E_{kin}$, and the thermal (thermal plus
kinetic energy), inside the virial volume at $z=0$.
The results are reported in the 4th column of Tab.\ref{tab:clusters}.
The kinetic energy of each cluster has been computed after subtracting the velocity of the center of (total) mass from the 3--D velocity field.
This parameter provides an indication of the dynamical 
activity of a cluster associated to the overall amount of mergers (e.g. Tormen,
Bouchet \& White 1997; Vazza et al.2006).
However, this procedure may be affected by the presence of massive companions in major merger systems, since this would bias somewhat the estimate of the center of mass of the main cluster.  
Nevertheless, $X=E_{kin}/E_{tot}$
provides a meaningful estimate of {\it perturbed} system in a statistical
sense, while in some particular cases the value may be underestimated by the 
error involved in the center of mass estimate.
Indeed, Fig.\ref{fig:kin} shows that major merger systems statistically present a large value of this ratio,
with $X \geq 0.4$ in most of objects (with a maximum of $X \sim 0.8$ in 2 cases). 
In what follows, we will define as 
``merging systems'' those objects that present a  $X>0.4$ value, but did not 
experienced a major merger in their past, according to the previous definition.

According to the above classification scheme, our sample presents 4 ``relaxing''
objects (clusters with no strong merger for $z<1$), 6 ``merging'' objects (clusters at the early stage of a merger with a massive companion inside the AMR region) 
and 10 ``major merger'' objects (clusters with a  $M(t_{2})/M(t_{1})-1>1/3$
for $z<1$).

In Fig.\ref{fig:categories} we give a visual representation of 3 clusters 
representative of the above
dynamical classes, mapping in the plane of the image the
the maximum of gas density along the line of sight, the 
maximum temperature along the line of sight and the gas 
temperature for a slice crossing the center of each cluster.

The major merger cluster E1 (first row) presents a richness of gas 
substructures and the imprints of strong shock heating as a result
of the major merger event that the cluster experienced at  $z\sim 0.1$.
Also, the slice in gas temperature clearly unveils the presence of a cold
front in the cluster center, likely located in the core of the accreted
satellite.

The merging cluster E5A is characterized by a quite smooth distribution
of gas temperature, but with the asymmetric imprint of a large scale
accretion due to the in-falling companion, whose center is located
at the distance of $\sim 3 Mpc$ from the center of E5A.

The relaxing cluster E14 presents a quite regular distribution of gas
temperature and gas density, which approaches spherical symmetry in
the innermost region, and it is characterized only by minor accretion
episodes.

The complete visual survey of all the clusters in the sample
is reported in the Appendix (\ref{sec:appendix2}).

\subsection{The IRA-CINECA Archive}
\label{subsec:archive}

The results of our simulation, in terms both of raw outputs and
of post-processed data, have features that make them interesting to
a broad ``audience''. The high spatial resolution and numerical accuracy,
the large number of
available redshift outputs (time resolution), and the large number
of clusters (statistics) can be exploited
for different purposes, similar or
even different from the original objectives.

Consequently, most of the produced data have been openly published and
are available through the 
web portal http://data.cineca.it, in the
{\it IRA-CINECA Simulated Cluster Archive} section.
The direct outputs of the simulations are available, 
in a reconstructed (i.e. monolithic 3--D boxes of the various gas/DM fields) 
or in the native ENZO domain-decomposed
formats. In both cases, the files adopt the HDF5 standard.
For most of the clusters in the sample, the whole evolution of ENZO outputs
(one every two time-steps) is available from $z=1$ to $z=0$, making possible
a highly resolved
($\Delta t \sim 0.1Gyr$) time study of the clusters evolution in all
gas/DM fields.
In addition, the set of nested initial conditions for all clusters in the
sample is provided in the same repository, along with preliminary
reduction of data and processing pipelines.

Due to the complexity and the size of our data products, they
have been organized and managed by means of a specialized software:
iRODS (integrated Rule Oriented Data System - http://www.irods.org)
a data grid software system developed by the Data Intensive Cyber
Environments (DICE) research group at the University of North Carolina at
Chapel Hill and the Institute for Neural Computation (INC) at
the University of California, San Diego.

We have exploited a few of the most interesting features of iRODS. The data
have been organized according to a specialized directory hierarchy and
they have been described by a precise data model in terms
of associated meta-data. The details can be found in a dedicated paper
(Gheller et al. in preparation). Part of the meta-data is managed
directly by the iRODS integrated iCAT database (PostrgreSQL
based), while part of it requires the adoption of a simple complementary
relational database. This is due to the current limitations of iCAT, which, however
are expected to be overcome in the next iRODS release, leading to a more
homogeneous treatment of metadata. The metadata can be used both to retrieve
information about the corresponding data objects and to perform SQL based
search, which allows the users to explore efficiently and effectively the data archive,
rapidly discovering the data sets of interest.

The iRODS server runs on the CINECA's SP6 HPC system, where the data was
produced and is currently stored. This, in order to avoid expensive and potentially
unsafe (for data integrity)
data transfers across different storage devices. The native support of iRODS for
multistreaming data transfer protocols (e.g. GridFTP) will be
exploited to deploy high performance download services, necessary to
move huge data objects. Finally, the possibility of federating geographically
distributed data servers, may, in the future, be exploited to mirror the data
and make its access even more effective.

At the time of the submission of the present paper, only a few of such services
are available, being still in a development or consolidation phase.
However, data are available and can be obtained on request,
following the instructions posted
on the web portal. 

\bigskip

The authors encourage the public access of data and the use of them for
original scientific research and mutual collaboration.
{\it If a paper is published with simulations produced in this project,
the authors should cite the present paper, and acknowledge the support
of the public archive at CINECA.}

\begin{figure*} 
\begin{center}
\includegraphics[width=0.48\textwidth]{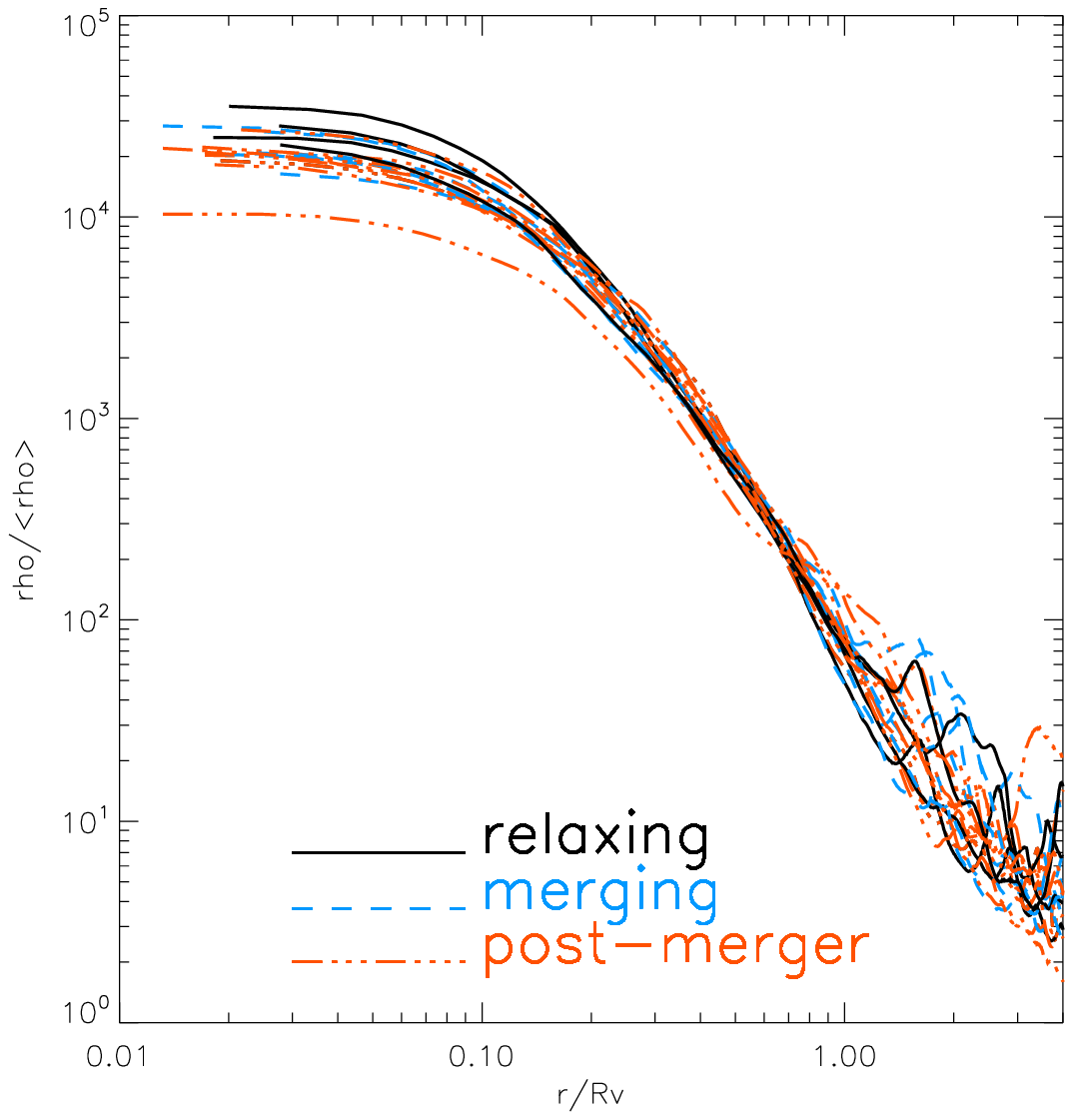}
\includegraphics[width=0.48\textwidth]{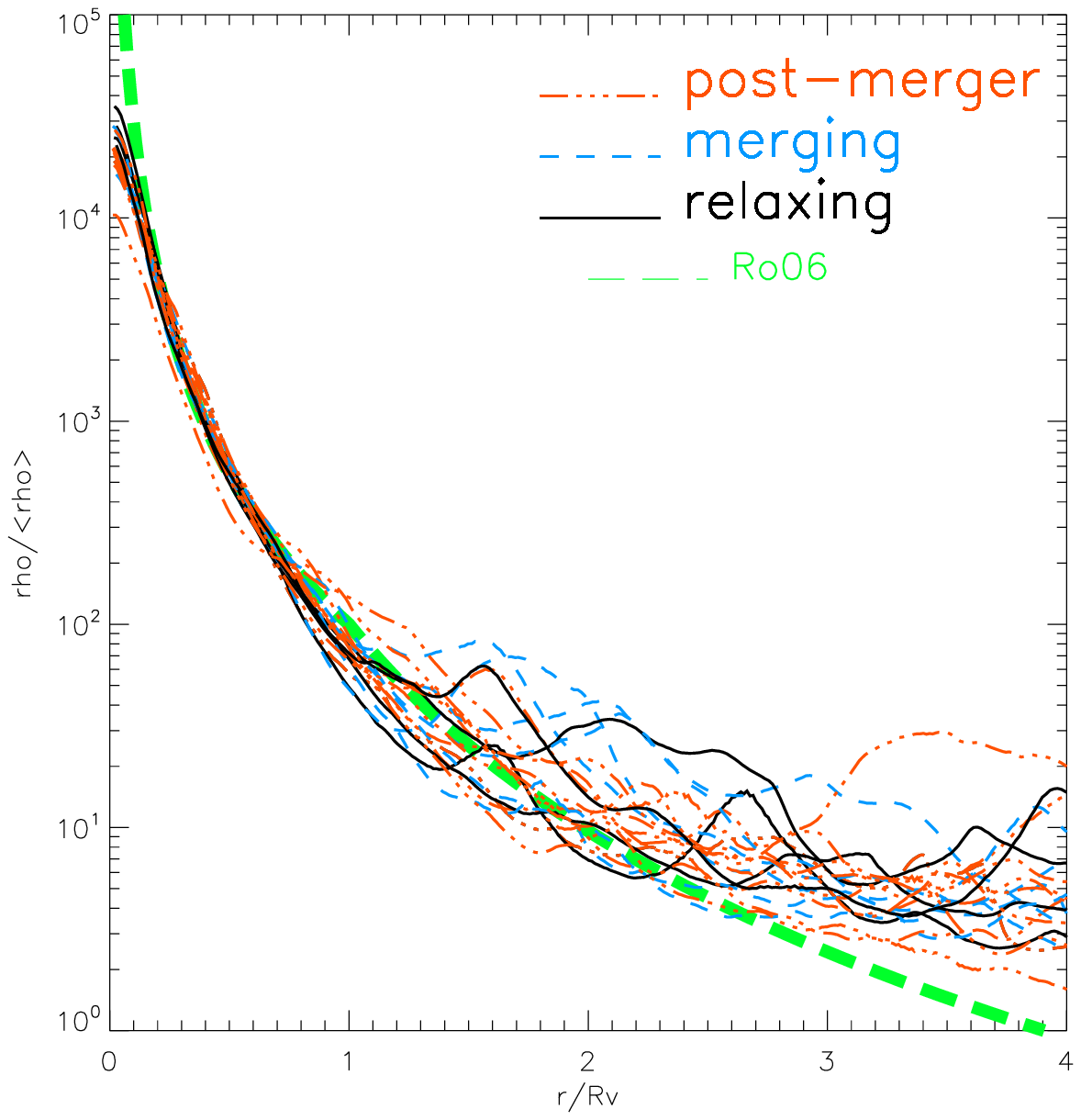}
\includegraphics[width=0.48\textwidth]{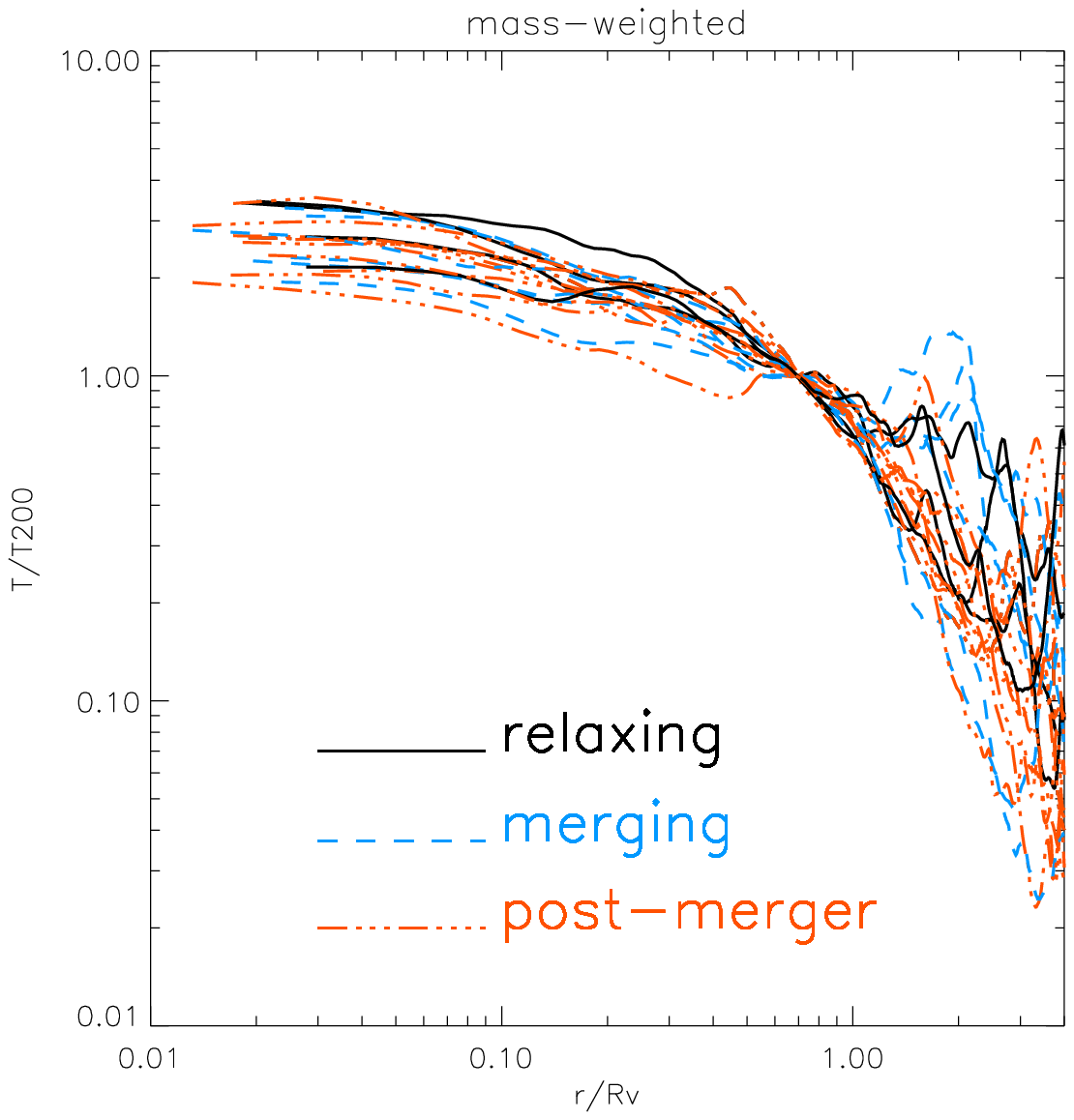}
\includegraphics[width=0.48\textwidth]{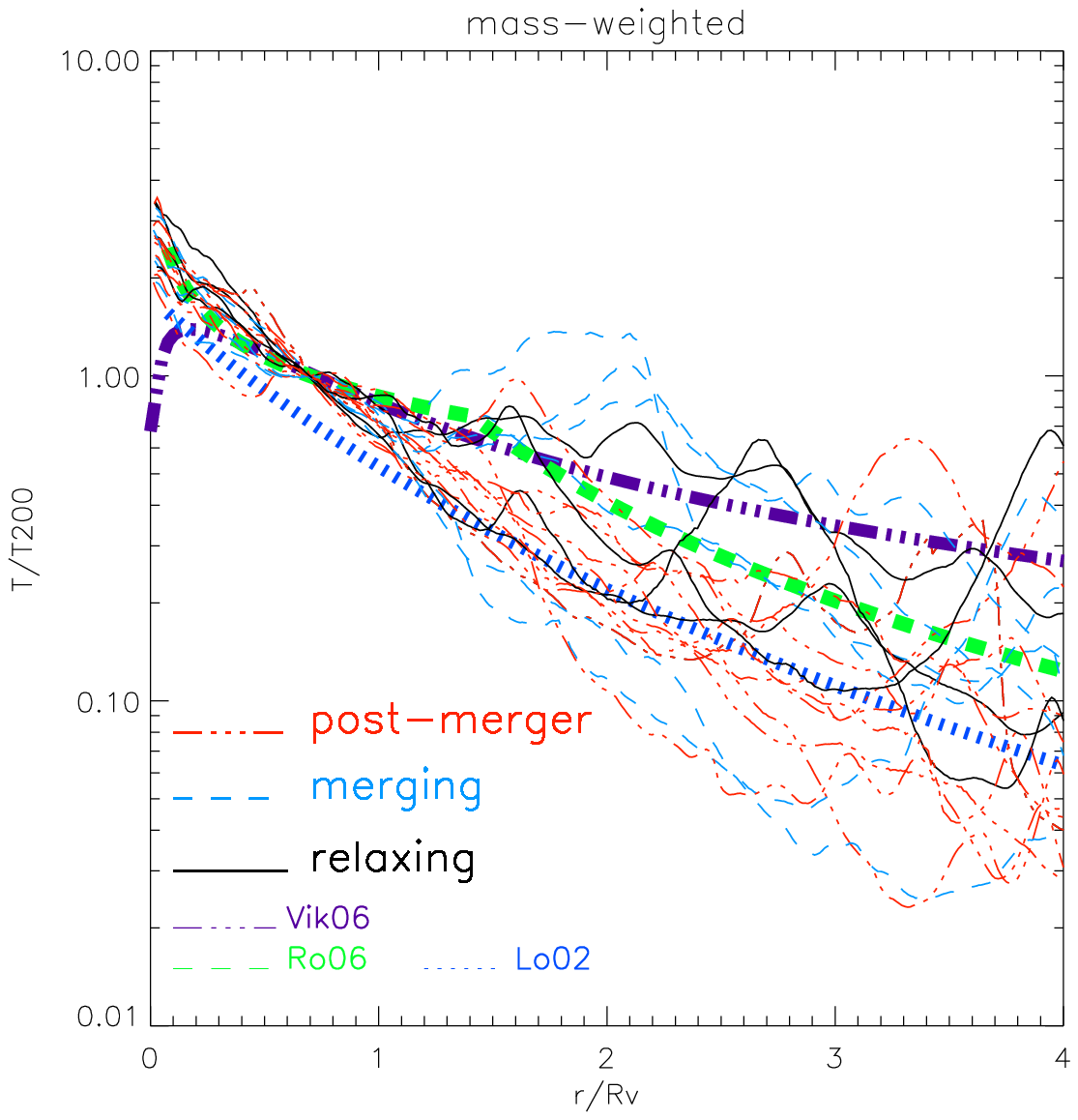}
\caption{Radial profiles of gas density (top panels) and gas temperature (bottom panels) 
for the clusters in the sample at $z=0$, with two different spacings for the X-variable. 
The different type of lines refer to clusters in a different dynamical state, according to 
Sec.\ref{subsec:clusters}. The additional lines are for the fit relations presented by: Roncarelli et al.(2006) as {\it green lines}, Vikhlinin et al.(2006), as {\it purple line} and Loken et al.(2002), {\it blue line}.}
\label{fig:dens_temp}
\end{center}
\end{figure*}

\section{Results}
\label{sec:results}
\subsection{Thermal properties: scaling laws}

\label{subsec:scaling}

It is well known that non-radiative simulations like those we report here,
present important differences compared to real clusters, and 
compared to cluster simulations with radiative cooling (e.g. Borgani et al.2008 for
a review). However, the most significant differences are found for halos
with  masses and temperature lower than those considered in this
work (e.g. for clusters with $T_{500}<1keV$), and 
therefore our clusters are expected to provide a viable representation 
of non cool-core systems.

As a first consistency check of the results produced by our cluster runs,
we computed the integrated values of $M_{500}$, $T_{500}$, $S_{500}$
and $Y_{X}$
for all clusters, where $M_{500}$ is the total (gas+DM) mass inside 
$r_{500}$ {\footnote{$r_{500}$ is defined as the radius enclosing a mean
cluster density of 500 times the critical density of the Universe; for
the assumed cosmological model this radius correspond to $\approx 0.5 R_{v}$.}}, $T_{500}$ and $S_{500}$ are the the mass weighted temperature and mass
weighted entropy (where the entropy is customary defined as $S=T/\rho^{2/3}$)
at the same radius, and $Y_{X}$ is the cluster total projected-pressure 
(Kravtsov, Vikhlinin \& Nagai 2006 ), measured
as $M_{gas,500}\cdot T_{500}$, where $M_{gas,500}$ is the gas mass inside $r_{500}$.

The panels in Fig.\ref{fig:scalings} show three meaningful scaling laws 
for galaxy clusters studies: the $T_{500}$ versus $M_{500}$, the 
$Y_{X}$ versus $M_{500}$ and the $T_{500}$ versus $S_{500}$ scaling
laws.

The ($T_{500}$,$M_{500}$) relation presented in the first panel in
Fig.\ref{fig:scalings} shows that our clusters follow the same scaling as 
in the self-similar model, $M_{500} \propto T_{500}^{3/2}$ (e.g. Kravtsov, Vikhlinin \& Nagai 2006), which is also consistent with  Chandra or XMM-Newton observations
of massive galaxy clusters (e.g. Vikhlinin et al.2005). The systematic bias
of $\sim 30$ per cent in the vertical direction is mainly due to the discrepancy between the 
real cluster mass measured in 3--D simulations and the derivation of 
mass from real clusters, under the hypothesis of hydrostatic equilibrium (e.g. Rasia et al.2006; Piffaretti \& Valdarnini 2008). 

The ($Y_{X}$,$M_{500}$) relation reported in the second panel in Fig.\ref{fig:scalings} is expected to
be the clusters scaling relation subject to the smallest intrinsic scatter (e.g. ``the perfect slope'', Kravtsov, Vikhlinin \& Nagai 2006). Indeed, our 
clusters follows the $M_{500} \propto Y_{X}^{3/5}$ scaling with an
extremely small scatter across one order of magnitude in $Y_{X}$.  The 
vertical systematics can be explained as in the previous case.

The ($T_{500}$,$S_{500}$) relation (third panel in Fig.\ref{fig:scalings}) 
for our clusters closely follows the self-similar prediction $S_{500} \propto T_{500}$ (e.g. Voit et al.2005). In the figure we show as a comparison the 
results obtained by Younger \& Bryan (2006) with an earlier version of the 
ENZO code (and in the case of no pre-heating of the ICM). The origin for the $\sim 10$ per cent bias in the entropy
at $r500$ is likely to be due to the slightly larger amount of entropy production (or entropy mixing) observed when standard mesh refinement is supplemented 
with the extra refinement triggered by using also velocity jumps, as shown in VGB10.

Our data proved to be compatible with the expected scaling relations, and this  suggests
that the global properties of our sample of clusters provide a meaningful
statistical representation of the most massive galaxy clusters in the Universe
at $z=0$.

\begin{figure} 
\includegraphics[width=0.48\textwidth]{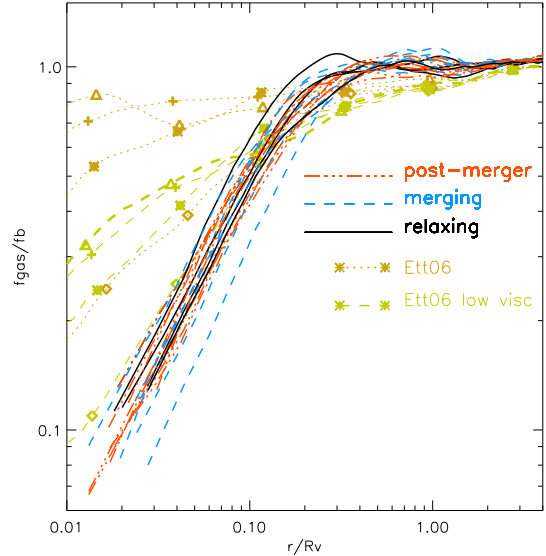}
\caption{Radial profiles of gas baryon fraction for the clusters in the sample at
$z=0$. The additional lines are for 4 galaxy clusters produced with GADGET2, with
standard viscosity (dotted lines) or a reduced viscosity scheme (dahsed lines), as
reported in Ettori et al.(2006).}
\label{fig:fbar}
\end{figure}

\begin{figure} 
\includegraphics[width=0.48\textwidth]{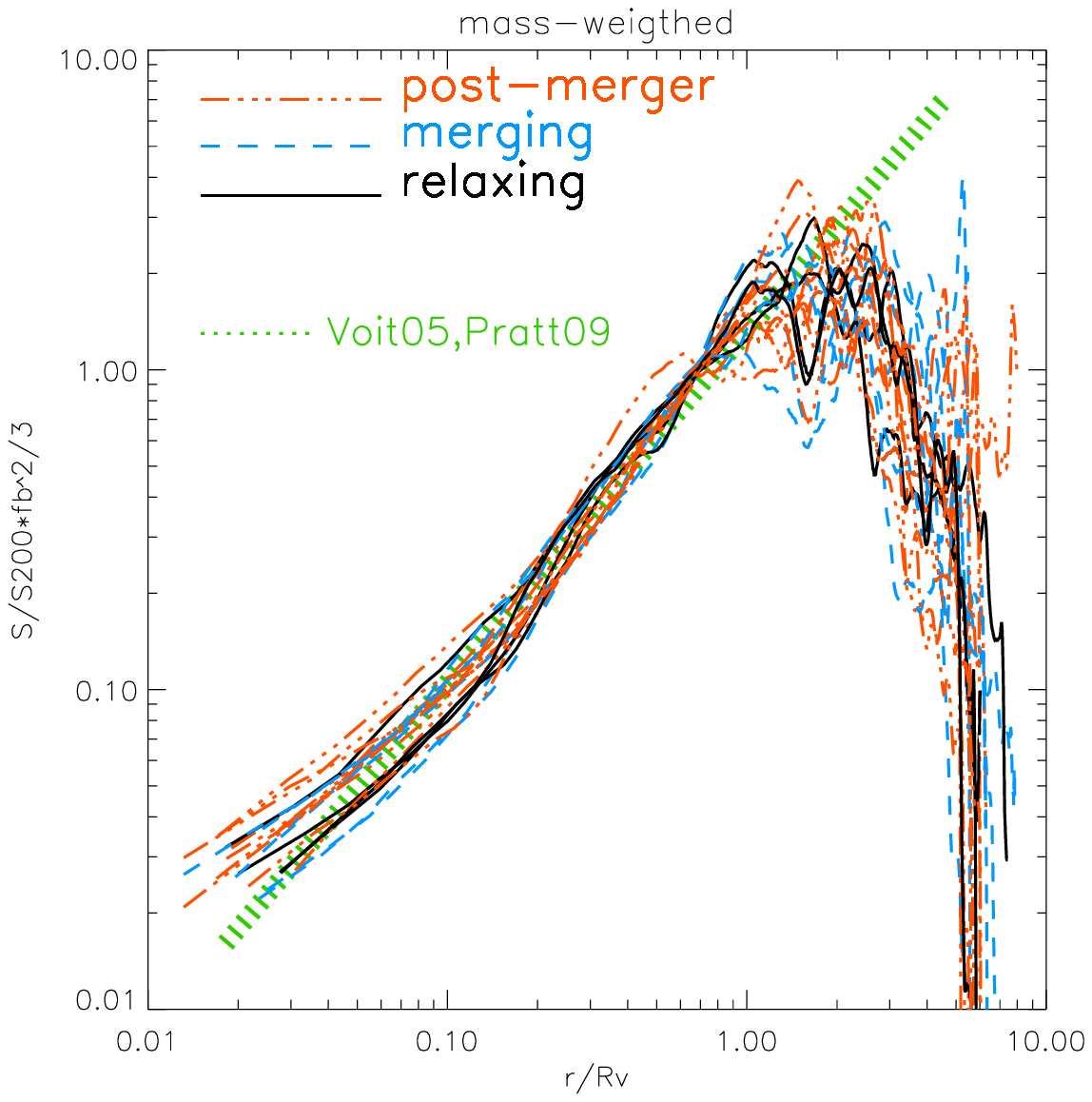}
\includegraphics[width=0.48\textwidth]{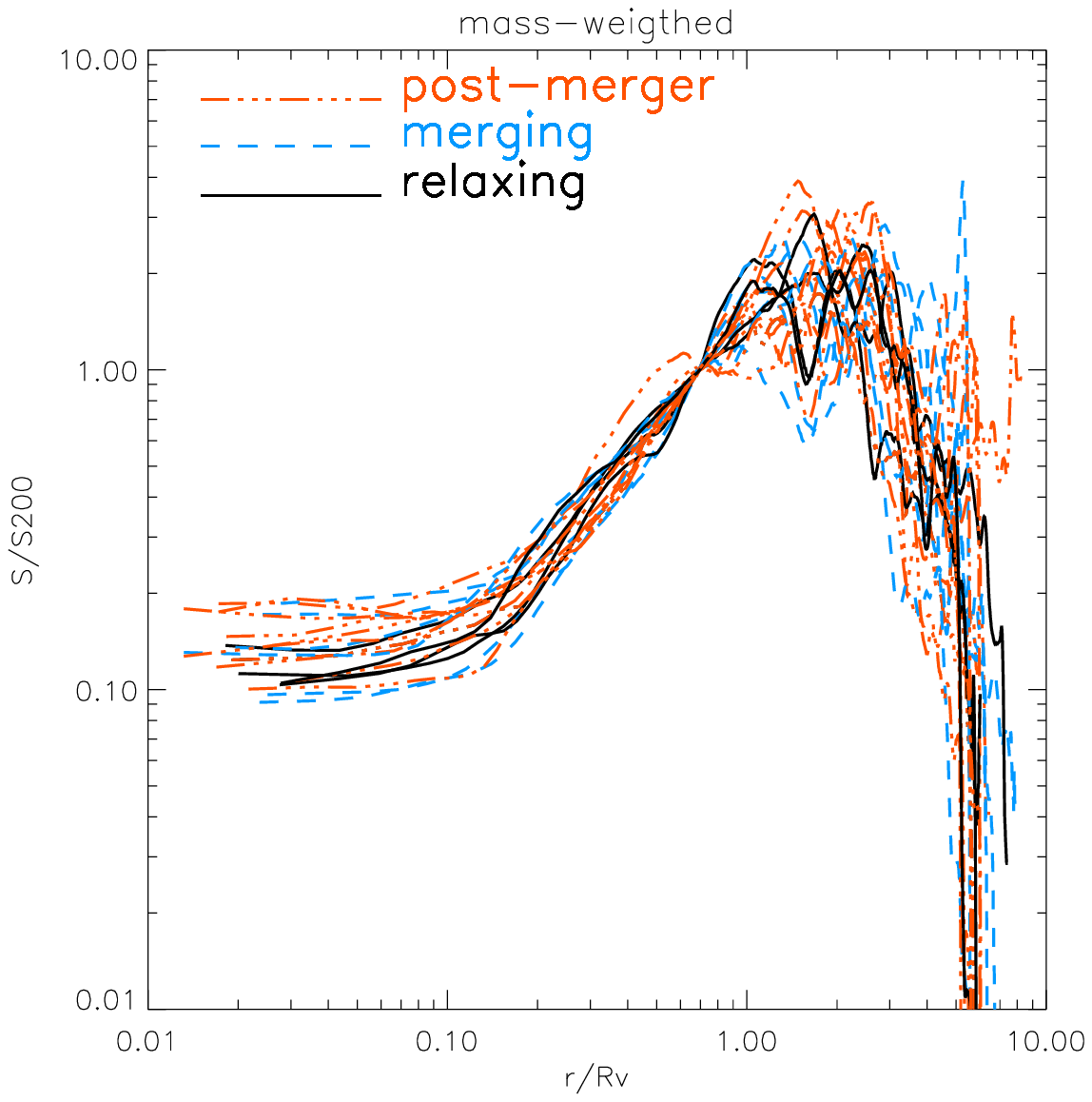}
\caption{{\it Top panel}: gas entropy radial profiles normalized by the baryon fraction within the
same radius, as suggested by Pratt et al.(2009). The additional line is the
best fit from Voit et al.(2005).{\it Bottom panel}: gas entropy profiles normalized by their value at $r_{500}$.} 
\label{fig:entr}
\end{figure}

\subsection{Thermal properties: radial profiles}
\label{subsec:profiles}

The high spatial resolution available in the peripheral
regions of our massive galaxy clusters ($\approx 25kpc/h$ up to a distance of 
$8-10 Mpc$ from the centers of clusters) 
provides a unique possibility of characterizing the
thermal properties of the clusters accretion regions using cosmological
numerical simulations with respect to what previously done in the
literature.

In the case of Smoothed Particles Hydrodynamics simulations
of galaxy clusters, the extremely high spatial 
resolution achieved in cluster cores (e.g. $\Delta \sim 5-10kpc$)
is quickly lost approaching the virial radius, due to the variable
smoothing length (e.g. $\Delta > 200kpc$), and makes it difficult
to obtain highly accurate spatial information for the cluster 
peripheral regions (e.g. Frenk et al.1999; O'Shea et al.2005); 
a similar effect is also present in Eulerian
simulations employing standard AMR techniques (e.g. Va09).  

Also the information from present X-ray observations of clusters is typically 
limited to the innermost cluster regions (e.g. $r<0.5 R_{v}$) and only very recently the SUZAKU satellite made possible to obtain
radial information out to larger radii for a few clusters
(e.g. George et al.2009; Reiprich et al.2009; Bautz et al.2009; Hoshino et al.2010; Kawaharada et a.2010).

In Fig.\ref{fig:dens_temp} we show the gas density and temperature profiles, 
centered on the center of total mass (gas+DM) of each object. All 
radii were normalized to 
the virial radius of each cluster, while values of
density and temperature were normalized to their values at $R_{200}$, 
$R_{200}$ being 
the radius inside which the mean density of the cluster is $200$ 
times the critical density of the Universe ($R_{200}\approx 0.7 R_{v}$).

The results are shown using both a logarithmic scale ({\it left panels}) 
to highlight the behavior in the innermost cluster region, and linear scale
({\it right panels}), in order to emphasize the large scale trends.
In most of  cases, the profiles present the evident imprints of 
massive and hot structures in the outer ($>2R_{v}$) regions, and in this
respect the classes of objects are characterized by a similar degree
of fluctuations. In the right panels of Fig.\ref{fig:dens_temp}, we additionally over-plot  as a comparison
the average behavior reported
in Roncarelli et al.(2006)
{\footnote{In Roncarelli et al.(2006) the
best fit of gas density and gas temperature profile is performed within
the range of $0.3 \leq r/R_{200} \leq 2.7$.}} 
for a sample of 7 galaxy clusters (4 with
total masses larger than $10^{15}M_{\odot}$) simulated with
GADGET2 (Springel 2005).  
The profiles from Roncarelli et al.(2006) fall within our cluster
statistics for $r<2 R_{v}$. However, since they focused on the
{\it smooth} gas component of the cluster matter (which implied 
a filtering out of gas substructures at all radii), the larger
mean density of our profiles at outer radii is easy explained by the contribution of gas cumps/filaments in our procedure.

We conclude that the average radial behavior of gas density and 
gas temperature for our clusters is consistent with
the other works in the literature which used complementary
numerical approaches to produce cosmological simulations of galaxy clusters.
On the other hand, our simulations provide an unprecedented look
at those external cluster regions.
Large fluctuations in gas temperature can be observed, and these
features  can be explained as the combined effect of having
more resolved shock waves, and more resolved accretion patterns
around in-falling satellites.
For instance, strong non-radiative shocks produce a post-shock density enhancement
of a factor 4, and an enhancement in temperature which scales with ${\it M}^{2}$
(${\it M}$ is the shock Mach number), and therefore 
resolved shock structures in our simulations produce sharp 
positive contribution to the mean radial density or temperature profiles.

Indeed, if we compare with previous results obtained by 
Loken et al.(2002) using a early version of the ENZO code with standard mesh
refinement (dotted blue line in the bottom right panel of Fig.\ref{fig:dens_temp}), we note that all radii our profiles show a slightly larger normalization and a larger amount of structures in temperature. This is consistent with the presence of high-temperature regions associated with strong and well-resolved pattern of accretion shocks in the outer atmosphere of our clusters, which are otherwise spread over larger distances if an overdensity-based refinement alone
is applied (e.g. Skillman et al.2008).

In Fig.\ref{fig:dens_temp} we also over-plot the mean temperature profile 
reported by Vikhlinin et al.(2006) from CHANDRA observations, and its
extrapolation at large scale. At large radii ($r>R_{v}$) the simulated clusters
show a steeper temperature profile compared to the extrapolated values
from the fitting formula of Vikhlinin et al.(2006):

\begin{equation}
T(r) \propto \frac{[(x/0.045)^{1.9}+0.45]}{[(x/0.045)^{1.9}+1]\cdot [1+(x/0.6)^{2}]^{-0.45}},
\end{equation}

where $x=r/r_{500}$.
Differently from the case of central cluster regions, 
the behavior of temperature at these density regimes is expected to be
mostly unaffected by any additional implementation of physical processes (radiative
cooling, star formation etc.), and therefore this should be a general
trend of Eulerian simulations. However, the contamination of in-falling 
substructures at all radii
leads to a large degree of scatter in our data. The presence
of particular direction of matter accretion (such as cosmic filaments) may 
also make the large scale distribution of temperature highly asymmetric, 
explaining
why the differences in the profiles of clusters with similar masses.
Interestingly enough, the presence of a filamentary structure of galaxies
was recently suggested by
Kawaharada et al.(2010) to explain the observed large scale
temperature anisotropy in a cluster Abell 1689, observed with SUZAKU. 

\bigskip

In Fig.\ref{fig:fbar} we show the radial profiles of the enclosed
baryon fraction, $f_{gas}$,  for every cluster in the sample. The baryon
fraction is normalized to the cosmic baryon fraction, $f_{b}$. As expected,
$f_{gas}/f_{b}$ reaches the value of $\approx 1$ at the virial radius of the clusters,
with a scatter $<10$ per cent when {\it relaxing}
and {\it merger} (or {\it merging}) clusters are compared.

The issue of the gas fraction distribution in galaxy clusters has
been extensively studied with numerical simulations in the last few
years, reporting small but systematic differences between Eulerian
and Lagrangian approaches (e.g. Borgani et al.2008 and references
therein). In particular, the gas fraction inside the virial
radius of clusters was found to be systematically higher ($\sim 10$ percent level)
in Eulerian AMR runs, compared to SPH runs (e.g. Ettori et al.2005; Kravtsov et al.2006).
 
We compare our profiles with 
the profiles of 4 galaxy clusters within a similar mass range, 
taken from Ettori et al.(2006). 
The dotted lines are for
GADGET2 simulations using the standard formulation for the 
artificial viscosity (e.g. Springel 2005), while the dashed
lines are for the re-simulations of the same clusters 
adopting a reduced viscosity
formulation (Dolag et al.2005). 
In all cases GADGET runs show larger baryon fraction within the
core region of clusters, although in the case of 
low-viscosity results the profiles approach the ENZO-AMR runs.
In ENZO PPM no artificial viscosity (beside the {\it numerical}
one) is present and due to our mesh refining procedure the
effect of shock waves and turbulent motions is maximized.
We can thus speculate that the basic difference in the 
inner profile of baryon fraction between SPH and grid
simulations 
is mainly due to the differences in mixing and stripping
of accreted satellites in the two schemes  (as early pointed 
out in Frenk et al.1999), which are both
affected by the presence of viscous forces in the
simulations. In runs without viscosity 
(PPM scheme or SPH with reduced artificial
viscosity) the stripping of in-falling sub clumps is more
efficient (e.g. ZuHone, Markevitch \& Johnson 2009) and
the stripped gas is distributed to larger radii compared
to runs with physical viscosity.

\bigskip

In the top panel of Fig.\ref{fig:entr} we show the entropy profiles, 
$S \equiv T/\rho^{2/3}$ (normalized by the value at $r_{200}$) for our clusters, multiplied by the
baryon fraction inside the same radius. As recently suggested by 
 Pratt et al.(2009), an universal profile may exist for
$S \cdot f_{gas}^{2/3}$, both in the case of cool-core and non cool-core
clusters. This may be explained by a scenario in which the feedback 
mechanism responsible for the increase of entropy in the innermost region of 
some galaxy cluster (e.g. AGN activity), also affects the radial distribution
of baryon gas at the same level. Therefore the product of the two
quantity levels out the two effects and the resulting profile is 
similar, regardless of the activity of a feedback mechanism.
Our data present a very good agreement with the reported correlation. 
This is also consistent
with the ($T_{500}$,$S_{500}$) scaling relation already discussed in Sec.\ref{subsec:scaling}.

In the bottom panel of Fig.\ref{fig:entr} we show the entropy profile
for each cluster (normalized to their value at $r_{200}$). 
All our clusters, almost independently with their dynamical
state, show a well-defined flattening of the entropy distribution
inside $<0.1 R_{v}$.

\bigskip

Eulerian simulations of galaxy clusters are generally known to convey higher
amount of entropy in the center of non-radiative galaxy clusters, by the 
combined effect of more efficient shock heating and mixing motions, compared to
SPH (Wadsley et al.2008; Mitchell et al.2009). 
Similar results
for re-simulations of galaxy clusters of intermediate masses ($M \leq 3 \cdot 10^{14} M_{\odot}$)
were reported in VGB10 using the same methods as in
the present work. 

Given the large resolution 
and the high number of clusters available in this sample, we confirm
this trend also in a statistical sense, reporting that {\it all} simulated
galaxy clusters present a well developed entropy core, with the 
size of $\sim 0.1 R_{v}$, with no evident relation with 
their dynamical state. 
This is of the order of $200-300kpc$, which is much larger than the 
softening length for the calculation of the gravitational force in the
PM scheme adopted by ENZO; therefore the entropy core found in our
simulations is not a numerical artifact.
Therefore, it would
be interesting to apply more complex models 
of the innermost cluster regions (e.g. by adopting cooling, AGN
feedback, etc) in order to measure the requested amount of 
extra-heating budget to quench catastrophic cooling, for the level
of entropy mixing modeled by these cluster simulations.

\begin{figure} 
\begin{center}
\includegraphics[height=0.99\textheight]{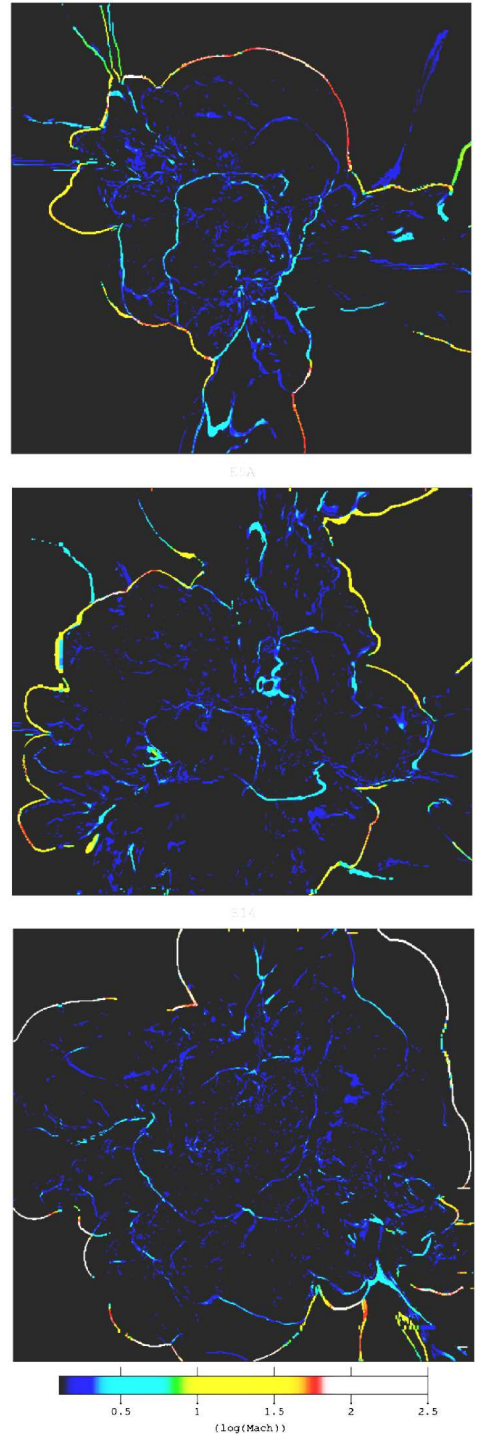}
\caption{Slices showing the Mach number of shocked cells for the same clusters and regions as in Fig.\ref{fig:categories}: E1 ({\it Top}), E5A ({\it center}) and E14 ({\it Bottom}).}
\label{fig:cut_shocks}
\end{center}
\end{figure}

\begin{figure} 
\begin{center}
\includegraphics[width=0.45\textwidth]{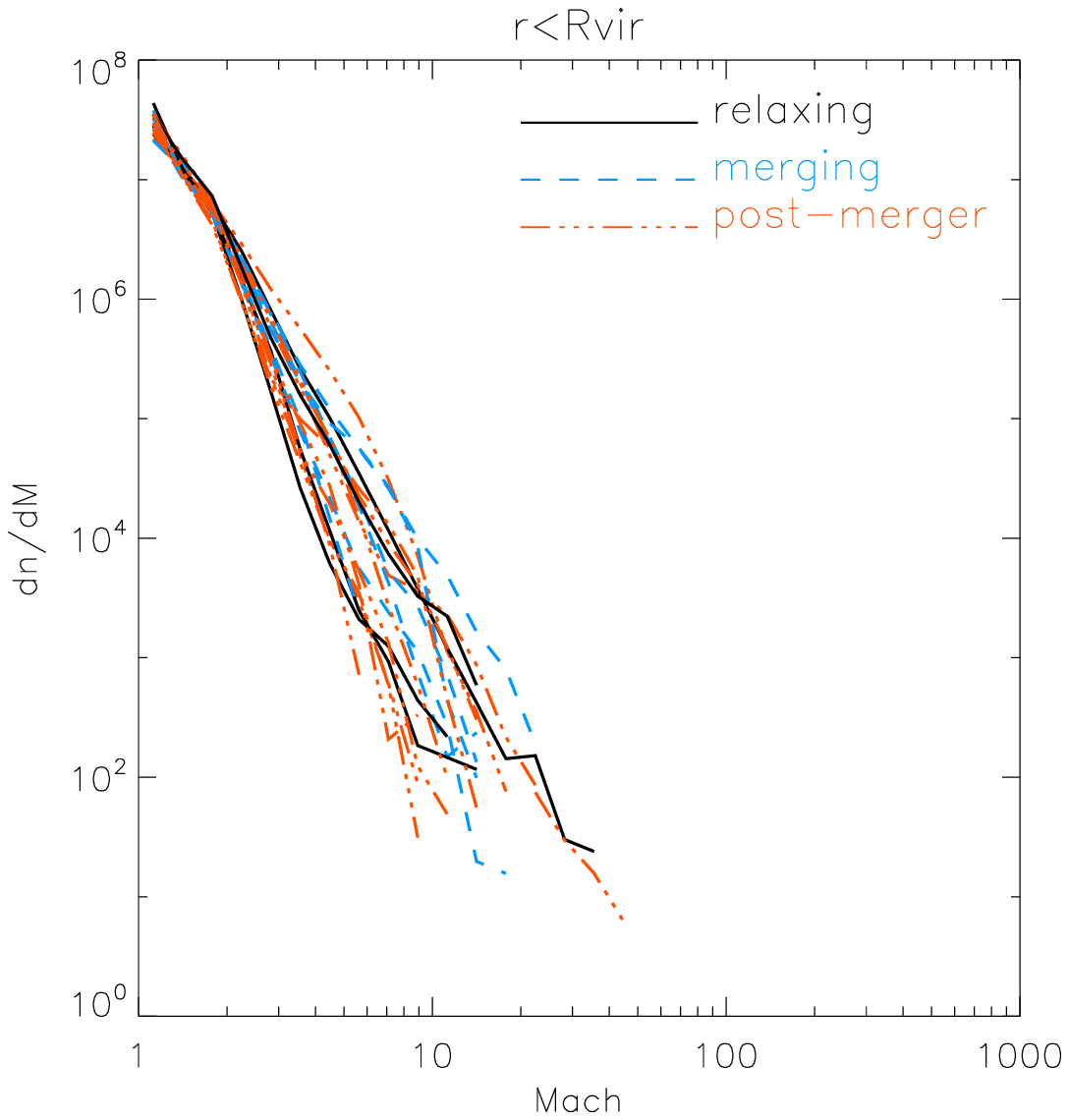}
\includegraphics[width=0.45\textwidth]{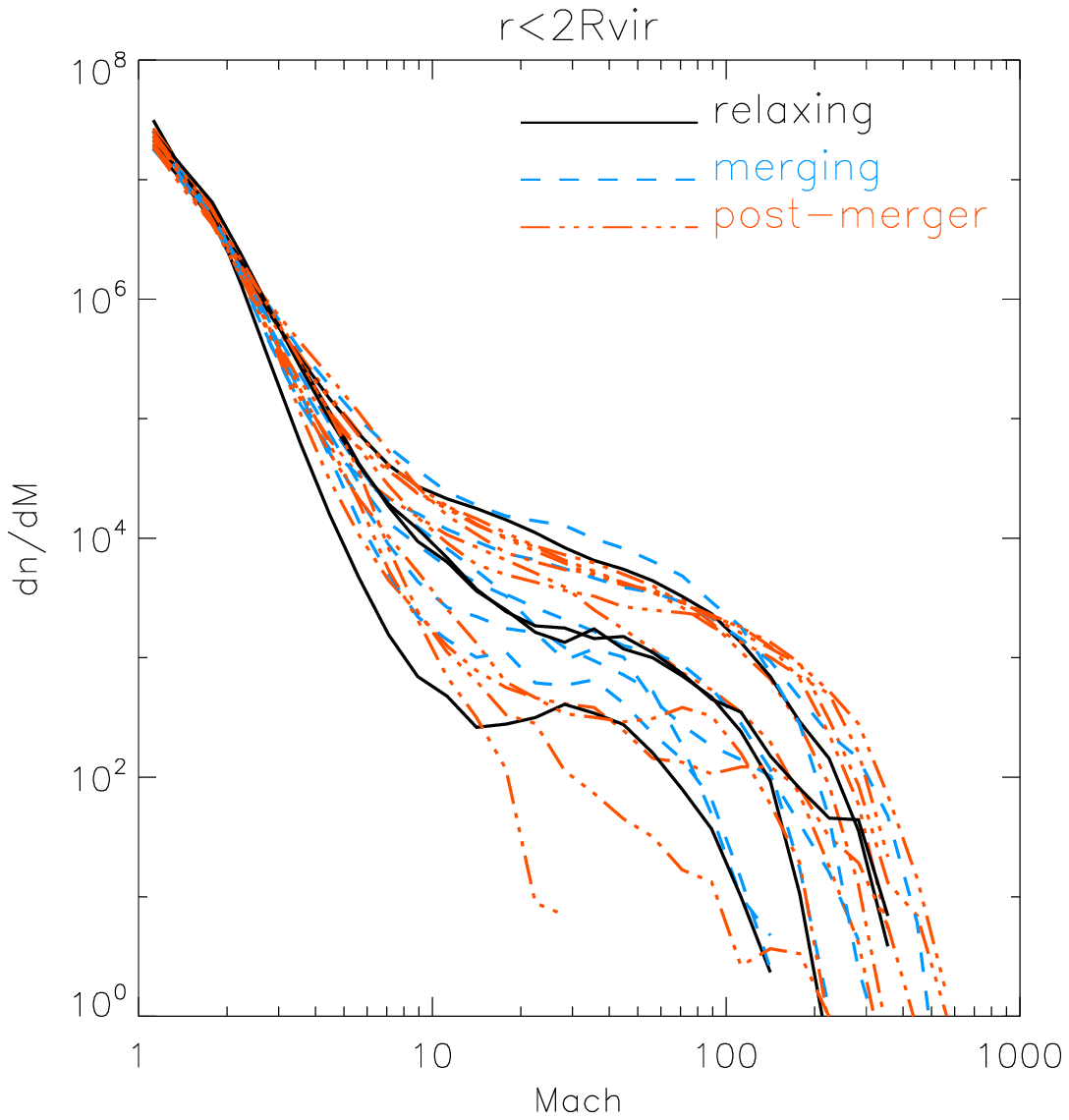}
\caption{Number distributions of shocked cells for $r<R_{v}$ and $r<2R_{v}$
volumes.}
\label{fig:dn}
\end{center}
\end{figure}

\begin{figure} 
\begin{center}
\includegraphics[width=0.45\textwidth]{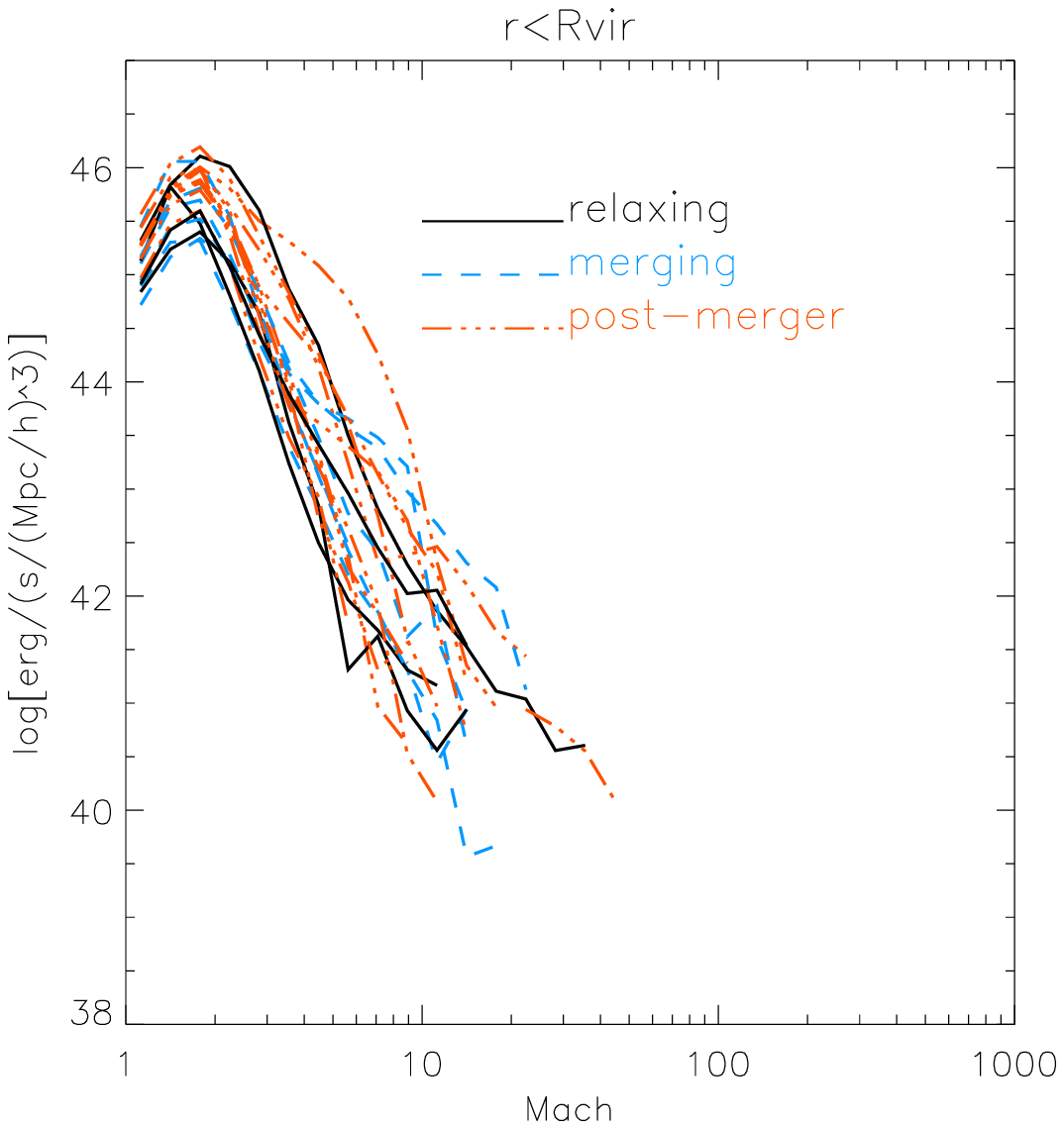}
\includegraphics[width=0.45\textwidth]{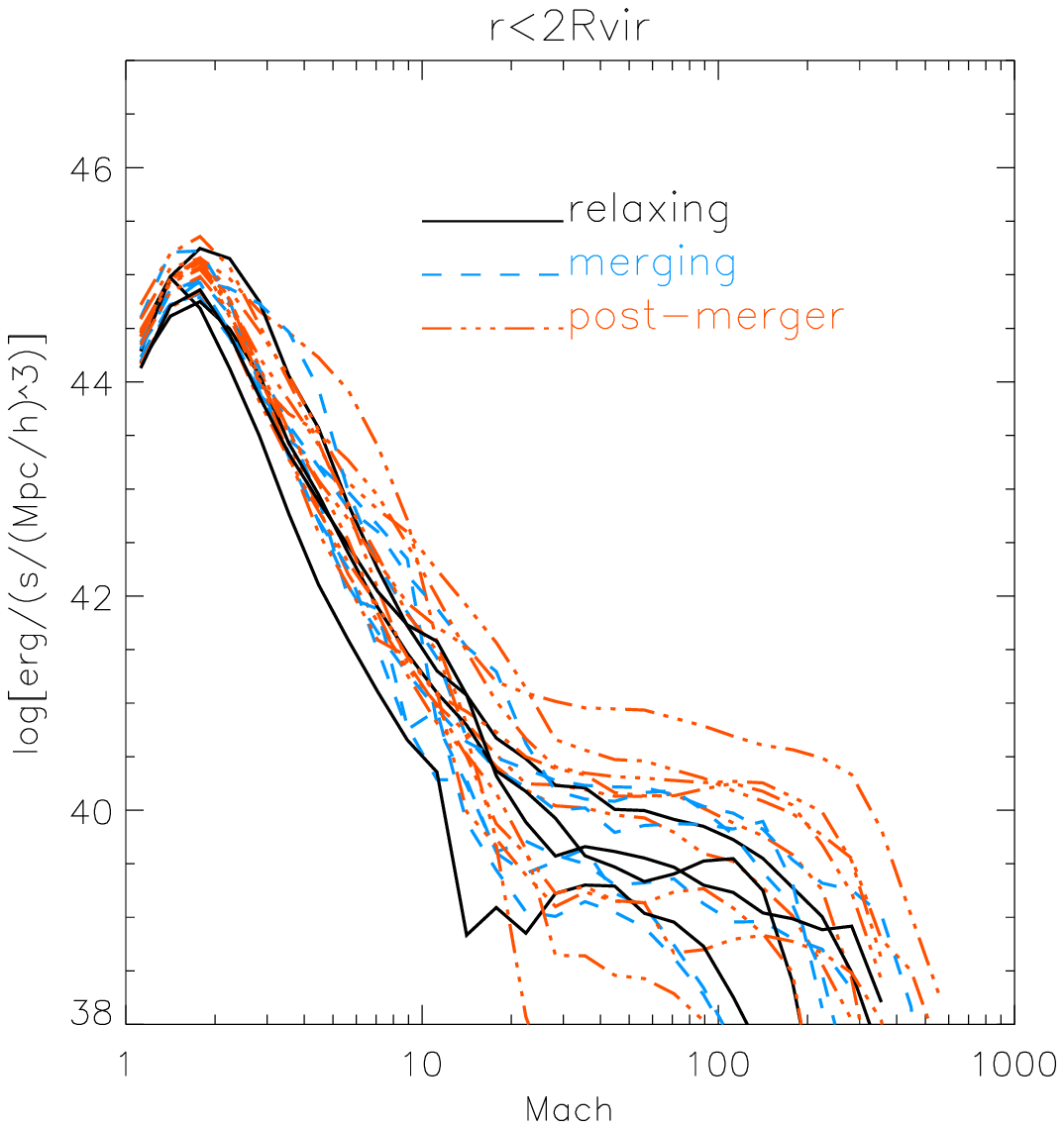}
\caption{Thermal energy flux distributions of shocked cells for $r<R_{v}$ and $r<2R_{v}$
volumes.}
\label{fig:flux}
\end{center}
\end{figure}

\begin{figure} 
\begin{center}
\includegraphics[width=0.45\textwidth]{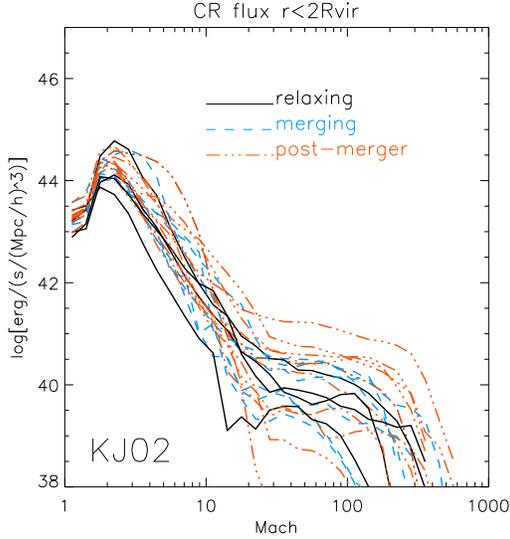}
\caption{Cosmic Rays energy flux distributions for $r<2R_{v}$, obtained 
with the injecetion model of  Kang \& Jones (2002).}
\label{fig:cr}
\end{center}
\end{figure}

\begin{figure} 
\begin{center}
\includegraphics[width=0.45\textwidth]{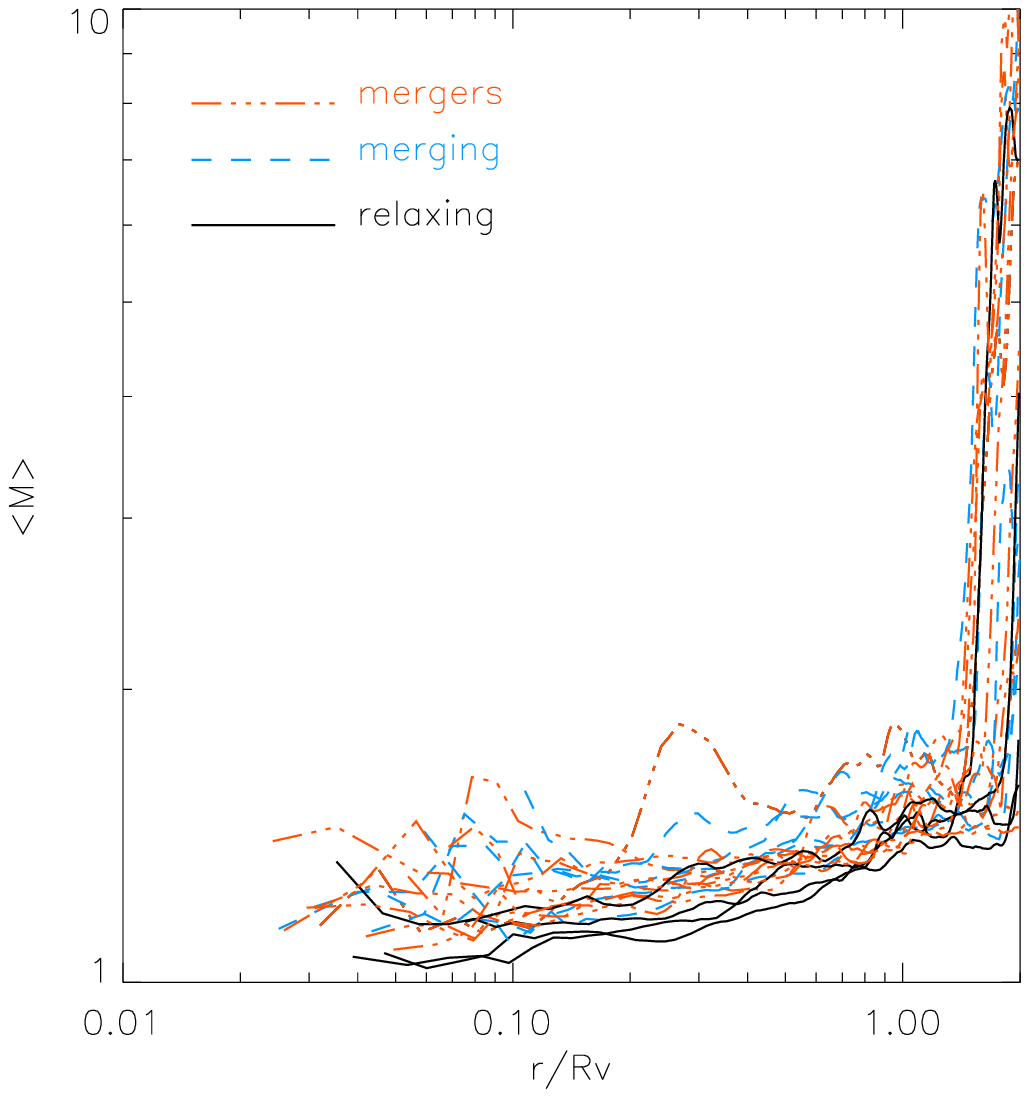}
\includegraphics[width=0.45\textwidth]{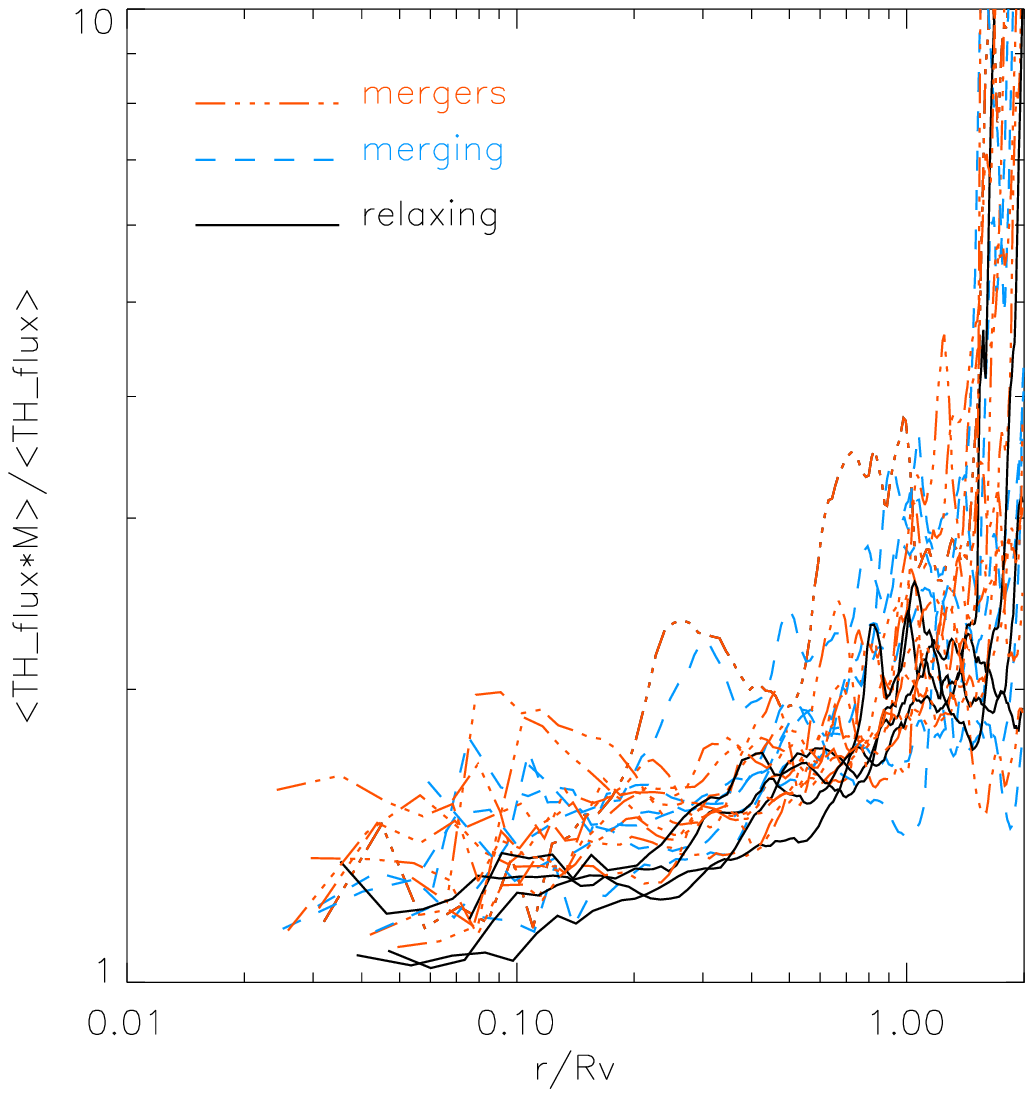}
\includegraphics[width=0.45\textwidth]{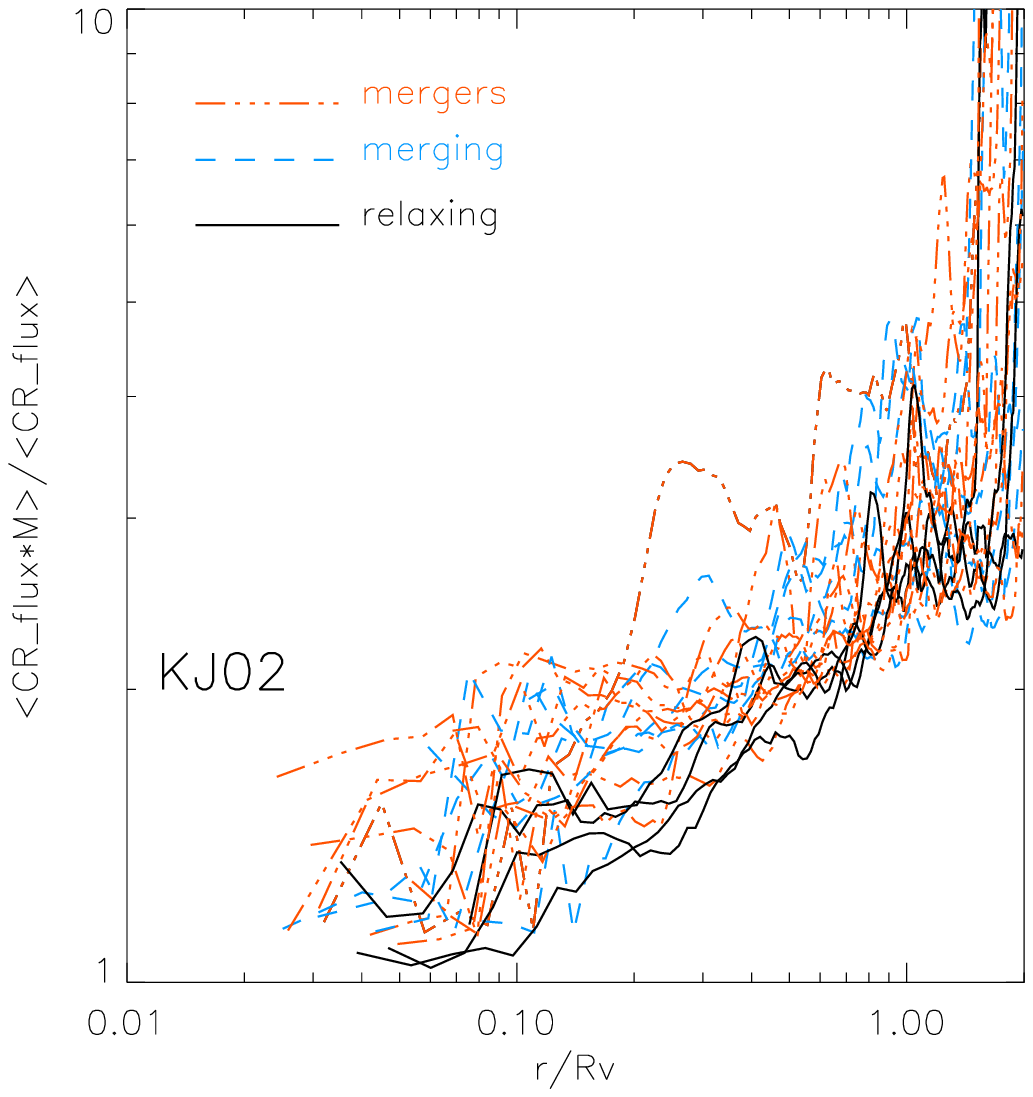}
\caption{Profiles of volume-weighted mean Mach number (top panel), thermal
flux weighted mean Mach number (middle panel), CR-flux weighted mean Mach 
using a Kang \& Jones (2002) injection model (lower panel).}
\label{fig:prof_mach}
\end{center}
\end{figure}

\subsection{Shock Waves.}
\label{subsec:shocks}

Our clusters sample is well suitable to 
study shocks statistics in the cluster formation region
even at large radii from the clusters center, 
since the refinement scheme presented in 
Va09 preserves the peak resolution of $25kpc/h$ on all shock 
features within the  AMR region. 

Observationally, merger shocks have been detected only in a few 
nearby X-ray bright galaxy clusters (Markevitch et 
al.2005; Markevitch 2006; Solovyeva et al.2008). They 
may be associated with single or double radio relics discovered
in a number of galaxy clusters (e.g. Roettgering et al.1997; 
Markevitch et al.2005; Bagchi et al.2006; Giacintucci et al.2008; 
Bonafede et al.2009). Shocks in large scale structures have 
been investigated in a number
of semi-analytical (Gabici \& Blasi 2003; Berrington \& Dermer 2003) 
and numerical works (Miniati et al.2001; Ryu et al
2003; Pfrommer et al.2007; Hoeft et al.2008; Skillman et al.2008, Vazza et al.2009; Molnar
et al.2009). 

We identified shocks with the same procedure presented in Vazza, Brunetti \& Gheller (2009), 
based on the analysis of velocity  jumps across close cells.
The preliminary 
selection of candidate shocked cells is made from the 
requirement that $\nabla \cdot {\rm v} < 0$;  the 
Mach number is finally evaluated 
from the inversion of: 

\begin{equation}
\Delta \rm{v} =\frac{3}{4}c_{s}\frac{1-\it{M}^{2}}{\it{M}^{2}},
\end{equation}

where $\Delta \rm{v}$ is the 1--D velocity jump between 3 cells across the candidate
shock, and $c_{s}$ is the sound speed of the cell with the minimum temperature. The full 3--D Mach number is then recovered as 
$M = (M_{x}^{2}+M_{y}^{2}+M_{z}^{2})^{1/2}$.

The panels in Fig.\ref{fig:cut_shocks} show the map of shock 
Mach numbers for the clusters E1, E5A and E14, taking the 
same slices as in the Right column of Fig.\ref{fig:categories}.
The differences in cluster dynamics translate also in significantly 
different large scale patterns of shock waves: a quite irregular and 
asymmetric pattern
of external shocks is observed in the case of the merging
cluster E5A. Also a few merger shocks can be found
in the virial region region of clusters E1 and of E5A, while
in the case of E14 only weak shocks can be found inside the
cluster core.

Following Vazza, Brunetti \& Gheller (2009) we calculated the volume distribution of shocks as
a function of $M$. This is shown in Fig.\ref{fig:dn}, for $r<R_{v}$ 
and $r<2R_{v}$.
Both distributions are steep, with an average slope of
$\alpha \sim -4 \div -5$  (with $\alpha = d\log N(M)/d\log M$) 
for $M<10$, consistently with results obtained with earlier
results obtained with ENZO at fixed grid resolution (Vazza, Brunetti \& Gheller 2009).
When the shocks distribution is computed for $r<2 R_{v}$, the differences
among clusters is found to be larger, especially for strong shocks with $M>10$.

This simply mirrors the scatter in the temperature distributions  
reported in Sec.\ref{subsec:profiles}.

The thermal energy flux across shocked cells is evaluated with

\begin{equation}
f_{th} = \delta(M) \cdot \rho M^{3} v^{3}_{s}/2,
\end{equation}

where $\rho$ is the pre-shock density and $\delta(M)$ is a 
monotonically increasing function of $M$ (e.g. Ryu et al.2003).

Figure \ref{fig:flux} shows the thermal energy flux distributions 
as  a function of the Mach number, for $r<R_{v}$ and $r<2R_{v}$.
The thermal energy flux for each cluster has been 
rescaled  assuming the volume of a sphere of radius $\sim R_{v}=3Mpc$ in order
to highlight the trend only due to cluster dynamics.  
Both distributions present a well defined peak of maximum thermalisation
at $M \approx 2$, and are very steep for all clusters:
$\alpha_{th} \approx - 5$ (with $\alpha_{th}$ taken as $f_{th}(M)M \propto M^{\alpha_{th}}$),
in agreement with results based on fixed grid resolution runs,
reported in Vazza, Brunetti \& Gheller (2009).

Results based on SPH (Pfrommer et al.2007) show significantly
flatter distributions ($\alpha_{th} \approx -3$ to $-4$), while 
in our sample this can be found only for a few 
clusters subject to violent merger events. 
Also in this case, the larger differences among our sample are found for $M>10$
shocks external to $R_{v}$, and usually $\sim 10$ times more energy is
processed
by these shocks in the case of post-merger systems, compared to the relaxing ones.

We apply also a simple recipe to estimate the efficiency of injection
of Cosmic Rays protons at shocks, with a standard application of the Diffusive
Shock Acceleration theory (e.g. Kang \& Jones 2002, hereafter KJ02).
The adopted injection efficiency is a function of the Mach number only:

\begin{equation}
f_{CR} = \eta (M) \cdot \rho M^{3} v^{3}_{s}/2;
\label{eq:fcr}
\end{equation}

where $\eta(M)$ is a monotonically increasing function of $M$, and
its numerical approximation can be found for instance in Kang et al.(2007).
 
Figure \ref{fig:cr} reports the distribution of CR energy flux adopting
the KJ02 injection model. The bulk of 
CR energy injection is achieved for $M \approx 2$, and only 
a few merging/post-merger
clusters show a broader peak of injection up to 
larger Mach number, $M \sim 4$. 
The maximum difference can be as high as $\sim 100$ in the CR energy
if relaxing and post-merger systems are compared for $M>4$ shocks.

Figure \ref{fig:prof_mach} presents the radial distribution
of mean Mach 
number, for the volume-weighted
average or for the weighting with the thermal and CR energy flux 
discussed above. 
The volume-weighted profiles are {\it extremely flat}, in 
agreement with previous studies (Vazza, Brunetti \& Gheller 2009), 
with just some strong imprints of internal
mergers shocks which increases the average value up to $M\sim 2$ in some post-merger
systems.
The  same is true for the thermal energy weighted profiles, while the profiles
become slightly steeper when CR energy flux is the weighting
quantity.
Interestingly enough, in all cases the occurrence of shocks larger 
than $M>2$ within $R_{v}$ is a rare 
event, which is qualitatively in agreement with the rare 
frequency of observed merger shocks in clusters.
In particular, from the inspection of Fig.\ref{fig:prof_mach}, we find that
only two clusters host strong shocks inside  
$r_{500} \approx R_{v}/2$, with
$M \approx 2.7$ (E1) and $M \approx 3.5$ (E2) respectively.

\begin{figure} 
\begin{center}
\includegraphics[width=0.45\textwidth]{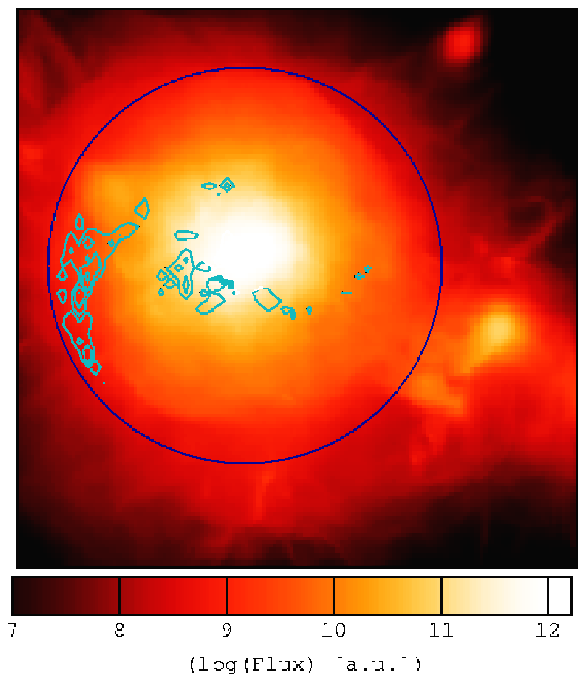}
\includegraphics[width=0.45\textwidth]{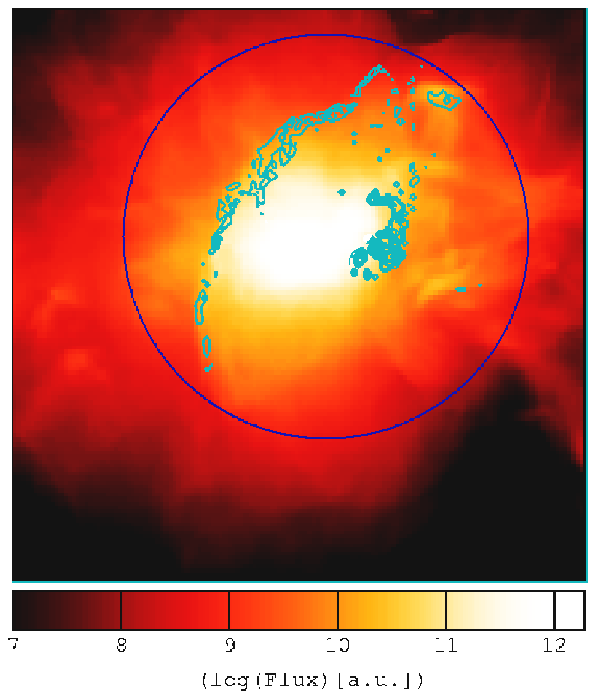}
\caption{Colors: projected bolometric X-ray emission for clusters E1 and
E2 (in arbitrary units); contours: CR-energy weighted maps of Mach number (only shocks with $M_{crw}>2.5$
are shown) for the same two clusters. The side of both images is $1.7 Mpc/h$,
the additional blue circles show the approximate location of $r_{500}$
for the two objects.}
\label{fig:ma_cr}
\end{center}
\end{figure}

In Fig.\ref{fig:ma_cr} we computed the projected bolometric X-ray
flux ($L_{bol} \sim \rho^{2} T^{0.5}$) for the innermost region
of the two clusters, and we additionally overlay the maps of CR-energy 
weighted mean Mach number ($M_{crw,ij}=<M_{xy} \cdot f_{CR,xy}>_{z}/<f_{CR,xy}>_{z}$, where the indices $x,y$ refer to the plane of the image, while the index $z$ runs along the line of sight) for a column of $2Mpc$ along the line of sight. 
In both cases, the shocks are $\sim 1Mpc$ wide and are located 
close to $r_{500}$.
It is intriguing that we found only two powerful $M>2.5$ shocks inside 
$r_{500}$ within a sample of 20 galaxy clusters. This ratio is roughly
similar to the ratio of clearly detected shocks in real clusters, which
are presently 3 out $\sim 30-40$ clusters imaged by CHANDRA
(e.g. Markevtich \& Vikhlinin 2007 and references therein).  
This is statistically compatible with the view that on average only
$1-2$ strong shocks cross the inner region of massive galaxy clusters
during their lifetime: $t_{life} \sim 5 Gyr$ for $M \sim 10^{15}M_{\odot}$
objects, while the crossing time of these strong ($M \sim 2-3$) 
shocks inside $r_{500}$ is of the order of $t_{cross} \sim 2 r_{500}/M c_{s} \sim 0.5 Gyr$,
which gives a chance of only $\sim 1/10$ to find a strong shocks crossing $r_{500}$
at a given time of observation.

In the Appendix (\ref{sec:appendix2}) we present the complete set of
projected CR energy flux for all clusters in the sample (showing
the estimated contribution from the accelerated particles of $E\approx GeV$),
with the overlaid bolometric X-ray luminosity for each cluster 
(Fig.\ref{fig:all_shocks}). The pattern
of projected CR fluxes in merging clusters tend to be very sharp, 
even if projected across
the whole cluster volume.

\begin{figure} 
\begin{center}
\includegraphics[width=0.45\textwidth]{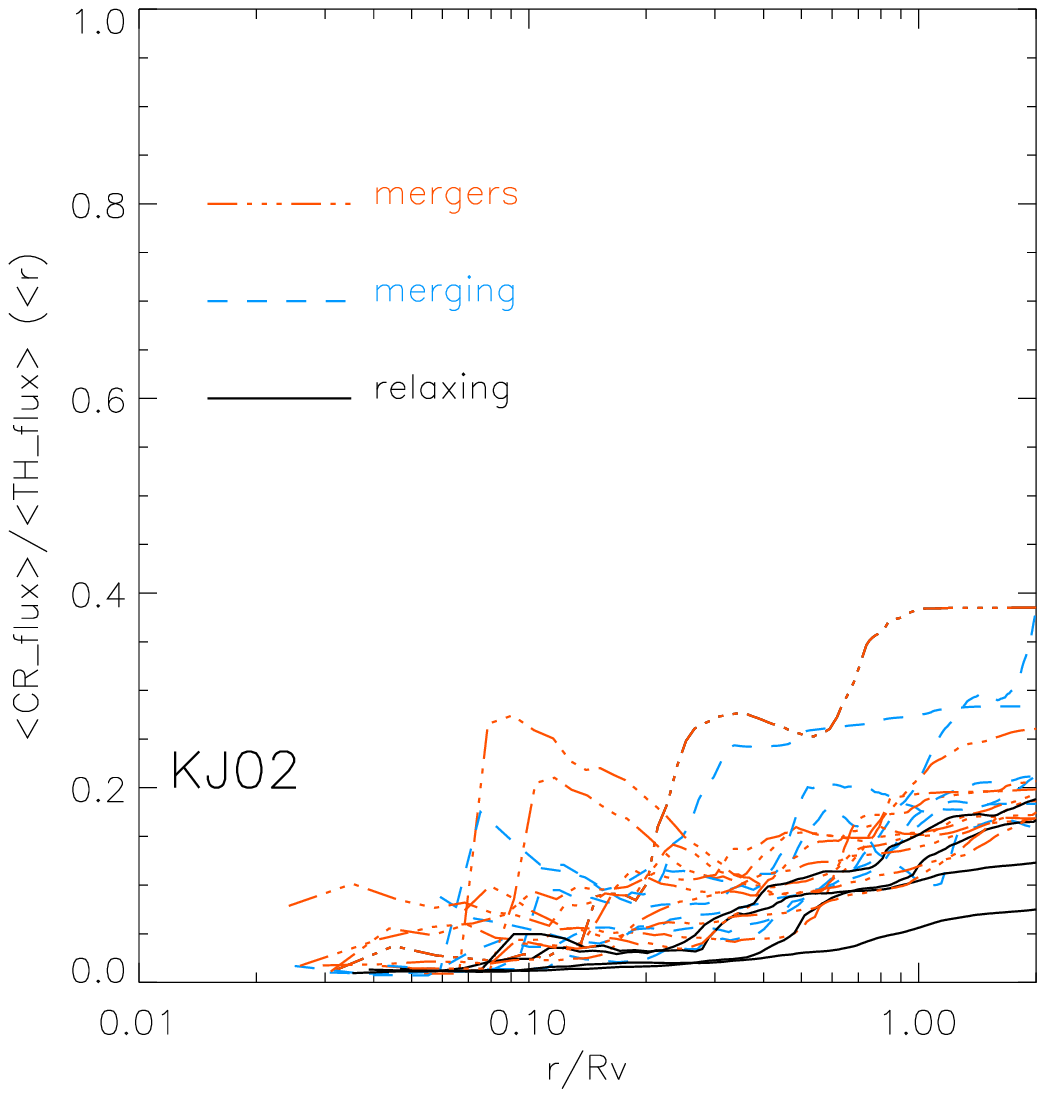}
\includegraphics[width=0.45\textwidth]{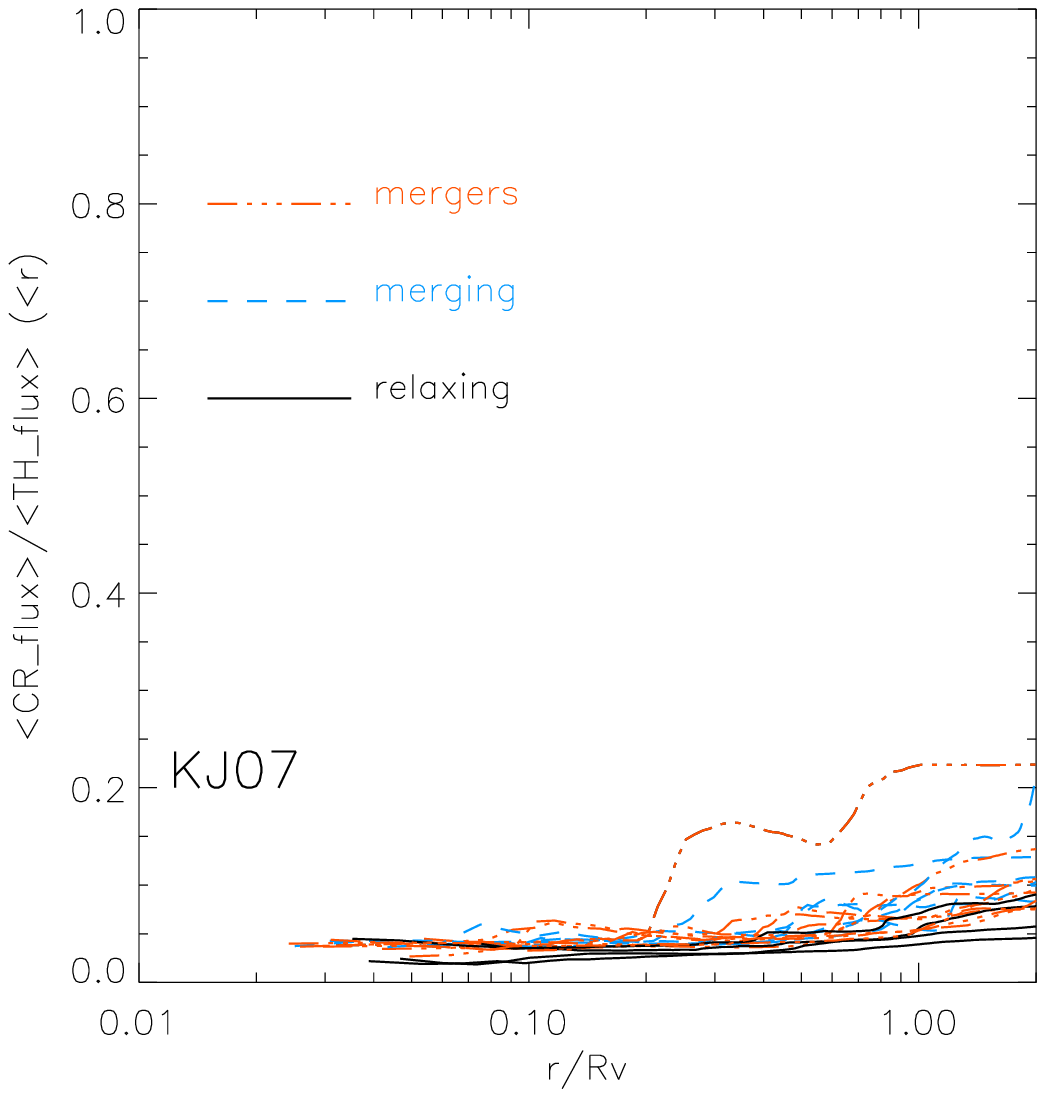}
\caption{Radial profiles for the integrated ratio between CR-flux and thermal energy flux.  
The top panel is for the Kang \& Jones (2002) injection model, while the bottom panel
is for the Kang \& Jones (2007) injection model.}
\label{fig:prof_ratio}
\end{center}
\end{figure}

Finally, we report in Fig.\ref{fig:prof_ratio} the integrated ratio between injected
CR energy and thermal flux inside a give radius, for the KJ02 model and also for 
a more recent model presented by the same authors (Kang \& Jones 2007).
In this second model, the effect of 
Alfv\'en wave drift and dissipation in the shock precursor are accounted
in a self consistent way, 
and this yields a value of $\eta(M)$ which is smaller than that
adopted by KJ02, at least for $M<20$. 
As a consequence the resulting distribution of
the energy flux dissipated in clusters by the acceleration of
CR as a function of the shock--Mach number is flatter than that obtained by adopting
KJ02, and the volume integrated injection efficiency are significantly
reduced (Kang et al.2007). 

In both cases the estimated ratio is $\sim 5$ per cent inside $0.2R_{v}$,
while inside $R_{v}$ the ratio is $\sim 20$ per cent for KJ02 and
$\sim 10$ per cent for Kang \& Jones (2007). Only one strong post-merger system 
shows the 
presence of a systematically larger ratio, $\sim 20-30$ percent inside $R_{v}$.
These results qualitatively support (with a much improved
statistical sample, and with $\sim 10$ better spatial resolution) 
the findings of Vazza, Brunetti \& Gheller (2009).

\section{Discussion and Conclusions}
\label{sec:conclusions}

In this paper, we presented a sample of 20 massive galaxy clusters 
in the range of total masses $6 \cdot 10^{14} M_{\odot} \leq M \leq 2 \cdot 10^{15} M_{\odot}$, extracted from large scale cosmological simulations and 
re-simulated with high mass resolution for the DM particles and
high spatial resolution for the gas component, up to $\sim 2-3 R_{v}$
from their centers.

We used the ENZO code with Adaptive Mesh Refinement, using a refinement
criterion based on gas/DM over-density and 1--D jumps in the velocity
field (as in Vazza et al.2009). 

With this approach, we obtained a statistical sample with unprecedentedly
large dynamical range within the virial volume of massive galaxy
clusters, which can be used to study in detail the thermal properties,
accretion phenomena and chaotic processes in the ICM over $2-3$
decades in spatial scale, for each cluster.

We presented the first exploratory statistic study of
this sample, showing the properties of gas density, gas temperature, gas entropy and baryon fraction for all clusters in our sample and the
radial profiles  in the range $0.01 \leq r/R_{v} \leq 3$ (Sections \ref{subsec:scaling}-\ref{subsec:profiles}).  The reported trends are in line with previous studies that used
complementary numerical techniques (e.g. SPH or standard AMR simulations),
however they  make possible to considerably extend the possibility of performing these
measurements at much larger radii, thanks to the high spatial resolution
in our simulations.

The additional mesh refinement scheme adopted in this work (based
on Vazza et al.2009) is explicitly designed to focus also on
shock features and chaotic motions leading to significant 1--D jumps
in the velocity field.
This allowed us to characterize the morphologies, frequency and
energy distributions of shock waves in these massive systems
(Sec.\ref{subsec:shocks} with unprecedented resolution
and to estimate their relative efficiency in accelerating
Cosmic Rays particles, by adopting two reference model of 
diffusive shock acceleration (Kang \& Jones 2002,07).
In agreement with previous studies based on much lower resolution
we confirm that the distribution of shock energy flux inside
clusters is extremely steep ($\alpha_{th} \approx - 4 \div 5$, with $f_{th}(M)M \propto M^{\alpha_{th}}$), that the peak
of the thermalisation at shocks is located at $M \approx 2$, and that the average 
 Mach number inside clusters is small, $M \sim 1.5$.

Only two clusters over 20 are interested by strong shocks inside
$R_{v}/2$ (at $z=0$), with $M \sim 2.7 $ and $\sim 3.5$ respectively.
The rarity of strong shocks found for $r<R_{v}/2$ is in line
with the statistics from X-ray observations of galaxy clusters.

We find that the injection rate of  Cosmic Rays is $\sim 5$ per cent of the 
thermalized energy within the clusters core, and 
$10-20$ per cent inside $R_{v}$, with a small dependence on the clusters
dynamical state. 
The behavior of shock surfaces in our simulations can be followed up to large
distances thanks to the resolution obtained with our AMR scheme, and potentially allows
us to describe with better detail the large scale shocks with respects to
precedent studies with SPH simulations (e.g. Pfrommer et al.2007; Hoeft et al.2008) or
grid simulations based on standard AMR (e.g. Skillman et al.2008).
We find energy flux distributions 
at shocks that are steeper than those obtained by Pfrommer
et al.2007 (with the energies flux reduced by $\sim 10$ times 
at $M \sim 10$). The morphologies of the shocks present
much sharper features (e.g. edges), even when projected  
across the cluster volume (see also Appendix \ref{sec:appendix2}).

At least part of this difference is due to 
differences in the effective resolution of shock waves inside and
outside clusters, which is preserved up to the maximum available resolution
by the AMR scheme in our runs (while it can strongly vary in SPH
simulations). Also, the differences in the thermal gas distribution at
large radii discussed in Sect.\ref{subsec:profiles} may 
play a role in giving difference in the properties of the shock
waves in the accretion
regions simulated by the two approaches.

In conclusion, we have presented a first look at a large
sample of massive galaxy clusters simulated with AMR techniques,
leading to an unprecedented level of spatial detail up to
large distances from the cluster centers.
This offers an important possibility to study the thermal
and non thermal properties of rich cluster of galaxies 
with a very large dynamical range in spatial scale, in a
fully cosmological framework.
A  public archive of data has been build
and made accessible via web (at the URL: http://data.cineca.it 
 under the {\it IRA-CINECA Simulated Cluster Archive} section), which
we encourage to access and use to produce further cluster studies and
to complement our results using other approaches.

 \section{acknowledgments}

We acknowledge partial 
support through grant ASI-INAF I/088/06/0 and PRIN INAF 2007/2008, and the 
usage of computational resources under the CINECA-INAF 2008-2010 agreement
and the 2009 Key Project ``Turbulence, shocks and cosmic rays electrons 
in massive galaxy clusters at high resolution''.
We thank S.Ettori for providing the data-points of the baryon fraction in Gadget simulated clusters. 
We thank A.Adamo, A.Bonafede, S.Cantalupo, R.Cassano, M.Meneghetti, M.Roncarelli and A. Vikhlinin  for 
useful discussions.

\bigskip

\appendix
\section{The reionization model}
\label{sec:appendix1}

We implemented in ENZO a run-time scheme to update the thermal energy of cosmic baryons, in order to reproduce with accuracy the effect of the a re-heating background due to stars and AGN activity. This is motivated by the fact that any re-ionization model
 with a gradual radiation turn-off can be well approximated by a suitably chosen sudden turn-on model (Hui \& Gnedin 1997). 
We adopted as fiducial model the re-ionization background by Haardt \& Madau (1996) spectrum supplemented with an X-ray Compton heating background from Madau \& Efstathiou (1999). The temperature structures of the re-heated cosmic gas across the whole AMR region cluster is very
well reproduced at all cosmic epochs, by imposing a temperature floor of $T_{0}=3 \cdot 10^{4}K$ in the redshift range of $2 \leq z \leq 7$. After this epoch,
we assume that the re-ionization background vanishes in a sharp transition.

Despite its simplicity (which turns into a less intense usage of memory and
computation) compared to the run-time re-ionization model implemented
in the public version of ENZO, we find that it works very well for the clusters
volume and its surroundings. 

\begin{figure} 
\includegraphics[width=0.48\textwidth]{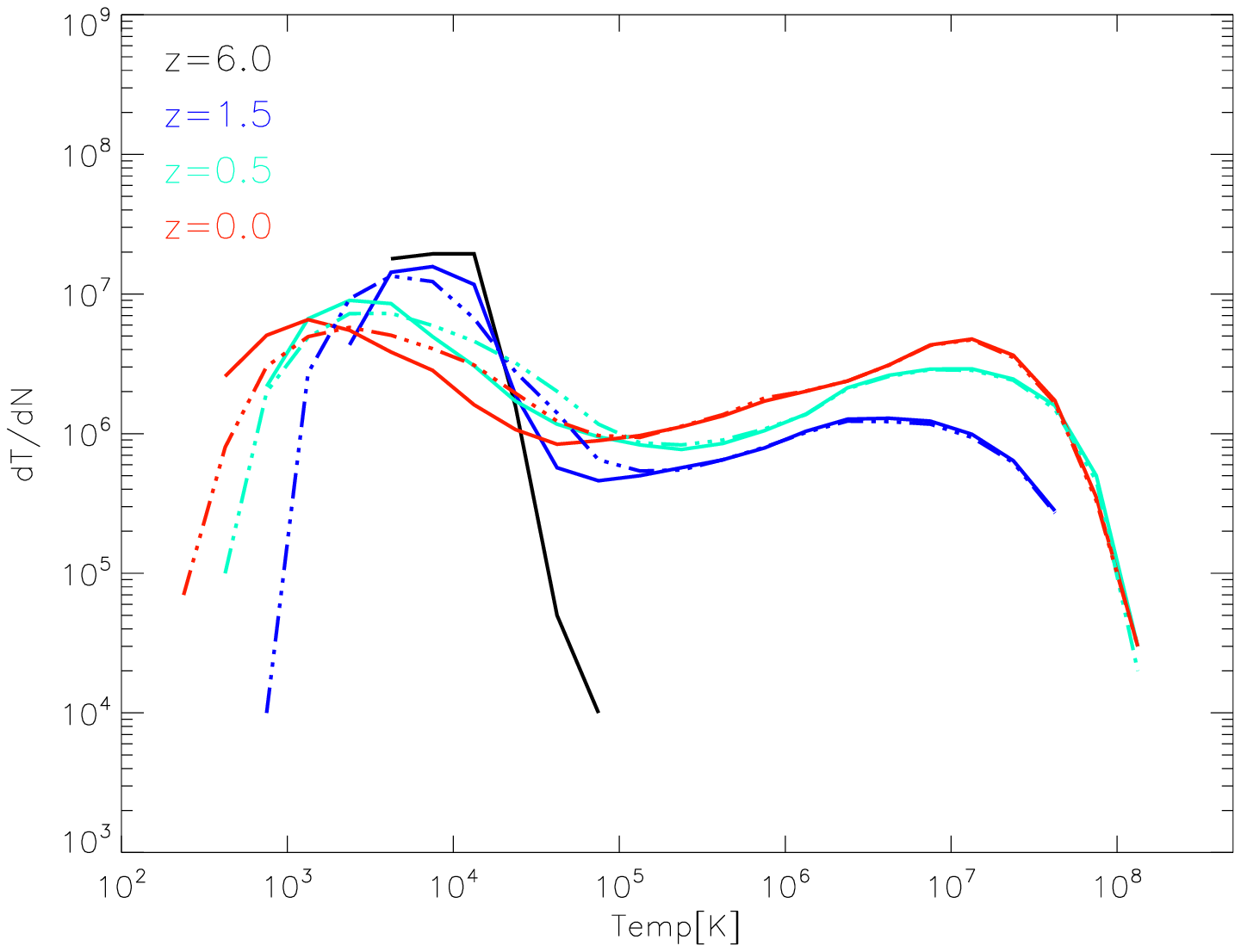}
\includegraphics[width=0.48\textwidth,height=0.4\textwidth]{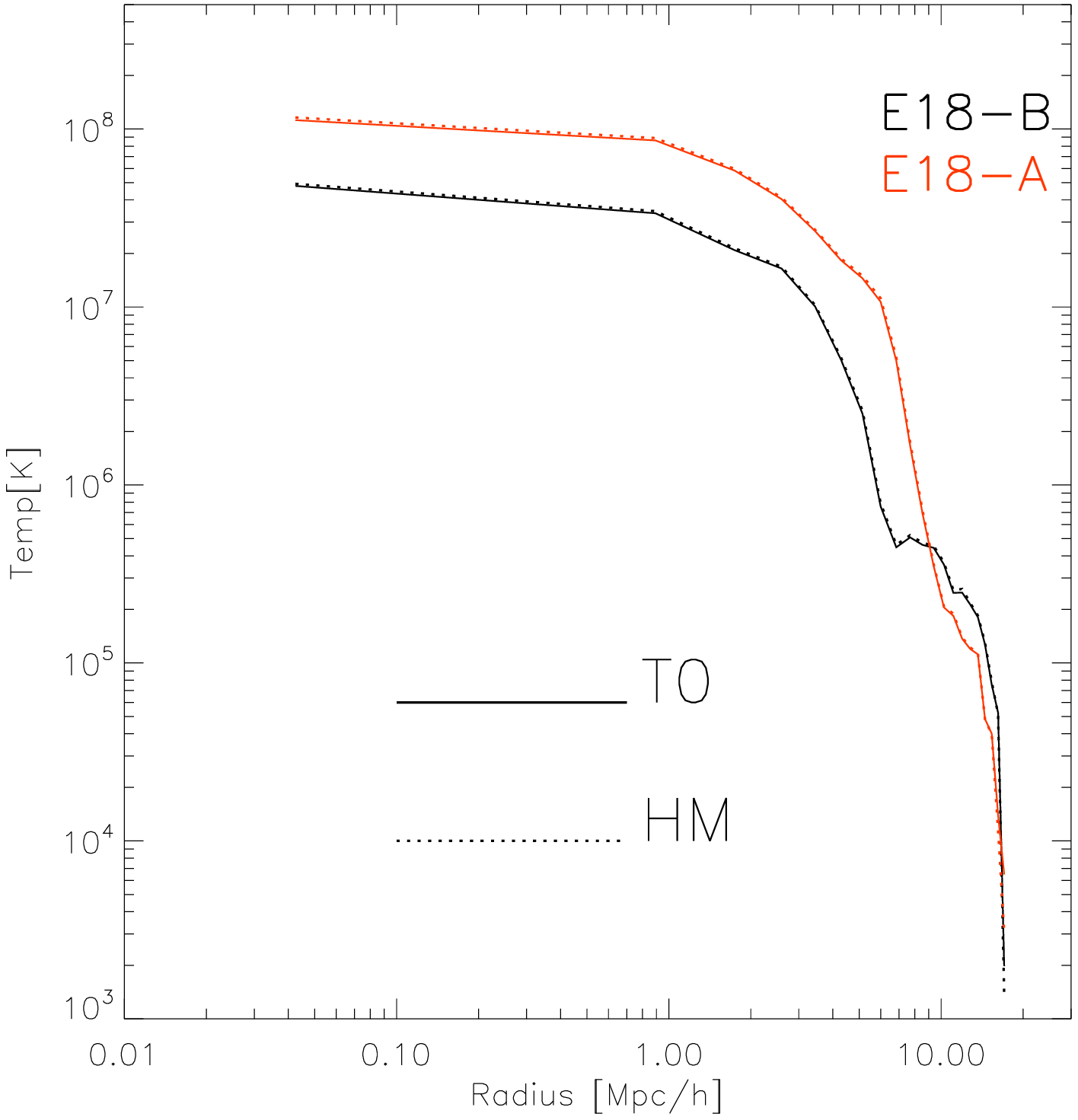}
\caption{{\it Top:} temperature distribution functions for the AMR region
at 4 redshifts, for the HM re-ionization model (dashed) and for the re-ionization model adopted in the present work (solid), for one of the simulated box of 187Mpc/h. {\it Bottom:} temperature profiles for the two most massive galaxy clusters formed in the same box as above. The solid lines show the dotted lines show the results for the HM re-ionization model, while the solid lines show the results fro the re-ionization model employed in this work.}
\label{fig:test1}
\end{figure}

In Fig.\ref{fig:test1} we report tests for the application of our modeling of re-ionization in ENZO (T0), comparing with the standard re-ionization scheme model
implemented in the public version of ENZO (HM).

Within the whole AMR region, the difference in the distribution is of the order of $\sim 10$ per cent at the lowest gas densities; if the radial temperature profile for the two clusters is computed, however, differences
are at the percent level for all radii up to $\sim 2R_{v}$ from the clusters
centers. Also the other thermodynamical properties of the clusters (e.g. gas density, gas entropy, etc) are recovered with a similar accuracy if comparing the two methods.

\section{Images of all clusters}
\label{sec:appendix2}

The visual inspection of projected maps or slices through
simulated clusters may provide additional and complementary information 
to more quantitative proxies, as those presented in Sec.\ref{subsec:clusters}.

In particular, the close inspection to projected maps taken from different
line of sight enhances the probability of finding physically meaningful 
similarity between simulated objected and real galaxy clusters observed in
X-ray (e.g. Sovloyeva et al.2007; Donnert et al.2009).

We encourage the interested readers in closely inspecting the whole
sample by using the public archive at  http://data.cineca.it/index.php,
and to report interesting similarities with real galaxy clusters.

\begin{figure*} 
\includegraphics[width=0.48\textwidth]{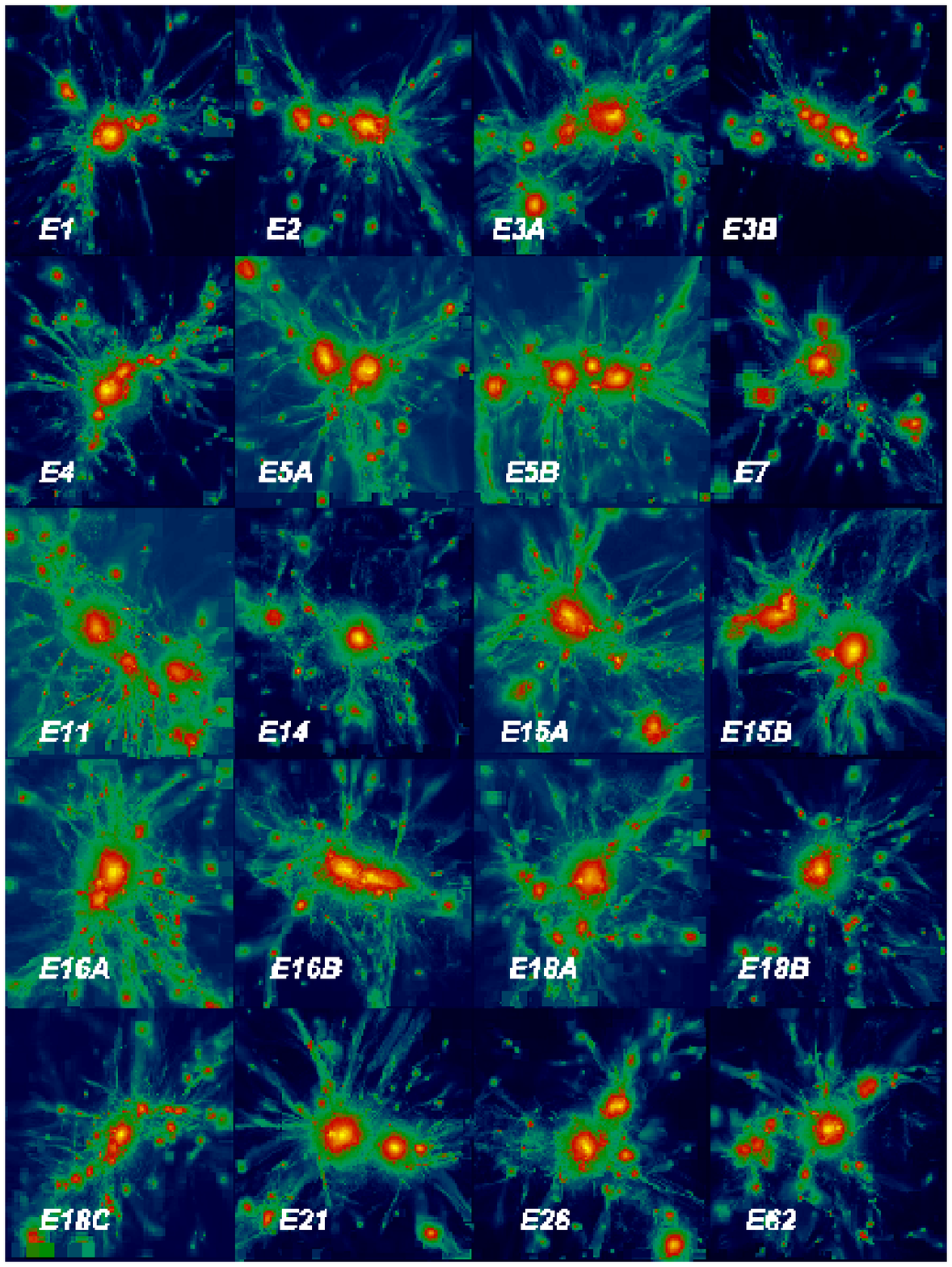}
\includegraphics[width=0.48\textwidth]{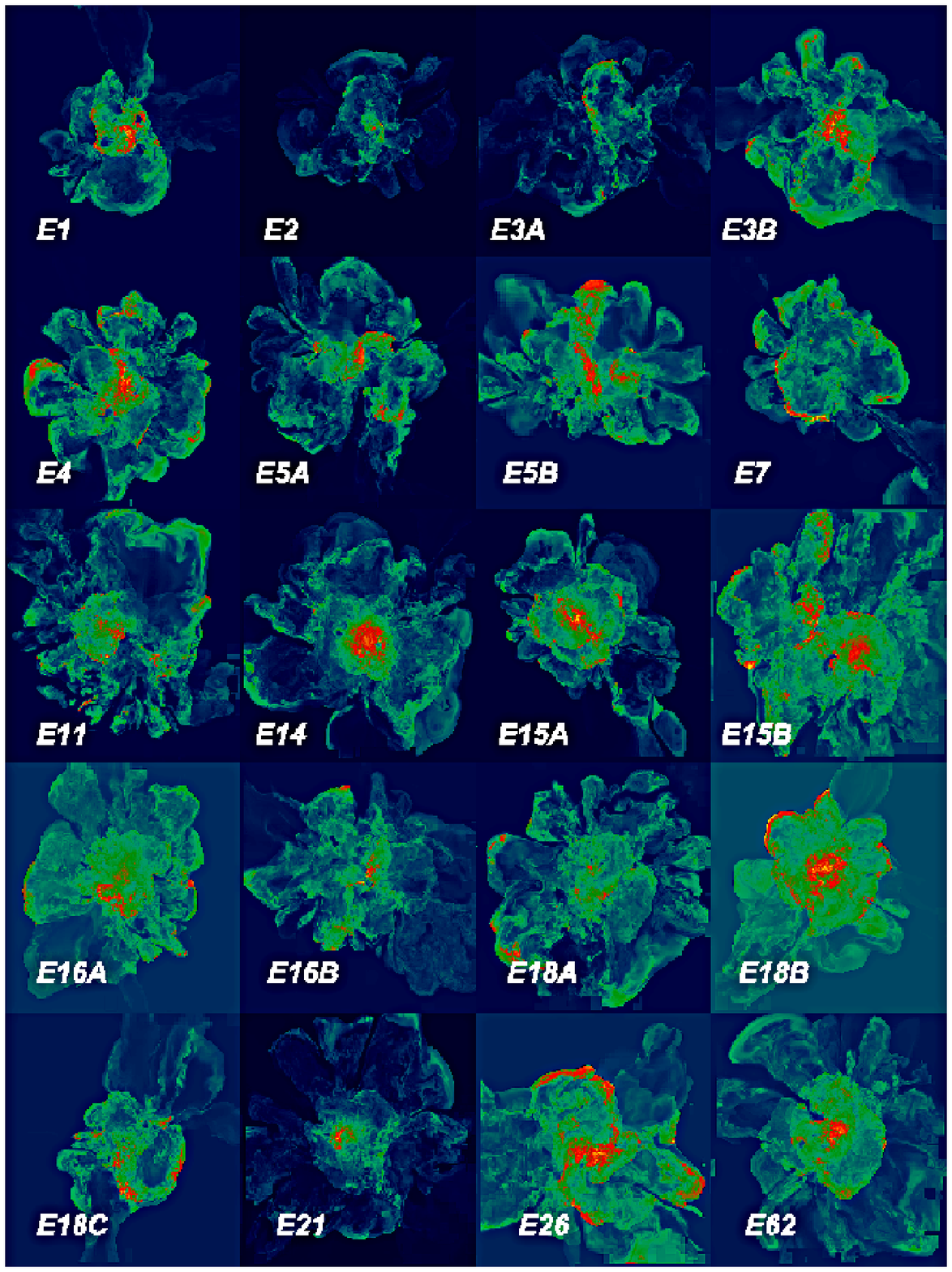}
\caption{Maximum gas density along the line of sight (left panel) and
cut in gas temperature (right panel) for all clusters in the sample.
The side of each image is $\approx 5 R_{v}$ of the enclosed galaxy
cluster.}
\label{fig:all}
\end{figure*}

The panels in Figures \ref{fig:all} show the projected 
gas density across the whole AMR region of all clusters simulated in the project (maximum along the line of sight) and slices in gas temperature, taken
through the center of mass of the same clusters. The size of all images
is rescaled to be $\approx 5 R_{v}$ of every object.

\begin{figure*} 
\begin{center}
\includegraphics[width=0.78\textwidth]{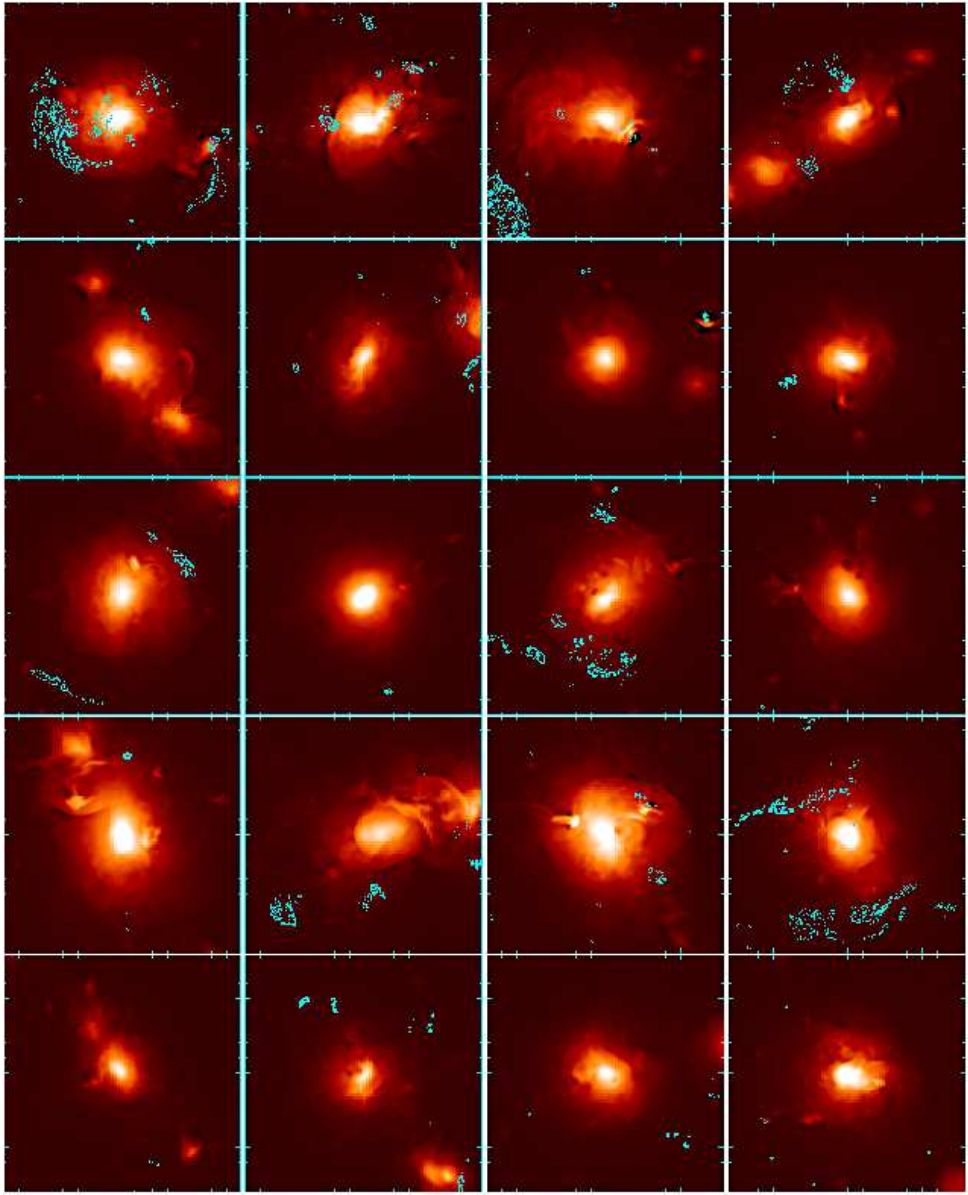}
\caption{Colors: projected X-ray bolometric luminosity for all
clusters in the sample (the color table is as in Fig.\ref{fig:ma_cr}).
Contours: energy flux for CR accelerated at shocks, for $E \approx 1GeV$
particles. The contours range for about 2 order of magnitude in flux, and 
they are spaced by $\Delta log(E)=0.2$.}
\label{fig:all_shocks}
\end{center}
\end{figure*}

The panels in Fig.\ref{fig:all_shocks} show the projected X-ray bolometric luminosity for all clusters in the sample (colors), and the energy flux along the line of sight for the Cosmic Rays particles accelerated at shocks (contours)
This gives a flavor of the patterns of shock waves associated with most of
the injection of CR particles, via Shock Diffusive Acceleration (as in Sec.\ref{subsec:shocks}). 
The CR particles flux was estimated by assuming a Kang \& Jones (2002)
injection model at each shocks, and by imposing a power law spectrum dependent
on the Mach number, $E(p) \propto p^{-s}$ (where $s=2(M^{2}+1)/(M^{2}-1)$). Only the flux associated to particles of $E\approx 1Gev$ is displayed
as isocontours (which span about 2 orders of magnitude in flux and are spaced in $\Delta log(E)=0.2$).
In at least three major merger systems (E1,E11 and E18B) we report the
evidence of couples of large scale, arc-shaped regions of intense CR acceleration of size $\sim Mpc$, which are reminiscent of doubles of radio-relics in real
galaxy clusters (e.g. Roettgering et al.1997; Bacghi et al.2005; Bonafede 
et al.2009).
 
\end{document}